\pdfoutput=1

\documentclass[%
superscriptaddress,
twocolumn,
 amsmath,amssymb,
floatfix
]{revtex4-1}

\usepackage{graphicx}
\usepackage{chemarr}
\usepackage{times,longtable}
\usepackage{hyperref}
\usepackage{version}
\usepackage{color}
\usepackage{soul}
\usepackage[version=3]{mhchem}

\usepackage{epsf,mathtools}
\usepackage[usenames,dvipsnames]{xcolor}

\graphicspath{{large/}}

\bibpunct{[}{]}{,}{n}{}{,}

\newcommand{\GL}{{\rm glyc}}
\newcommand{\D}{{\rm don}}
\newcommand{\A}{{\rm acc}}
\renewcommand{\L}{{\rm LDH}}
\newcommand{\LAC}{{\rm LAC}}
\newcommand{\M}{{\rm ox}}
\newcommand{\G}{{\rm G}}
\newcommand{\ATP}{{\rm ATP}}
\newcommand{\HEX}{{\rm HEX}}
\newcommand{\PDH}{{\rm PDH}}
\newcommand{\LDH}{{\rm LDH}}
\newcommand{\GLUN}{{\rm GLUN}}
\newcommand{\atpmin}{f_{\ATP}^{\min}}

\newcommand{\rev}[1]{\textcolor{black}{#1}}

\usepackage{chemarr}
\usepackage{epsf,mathtools}
\usepackage{graphicx}
\usepackage{dcolumn}
\usepackage{bm}

\begin{document}


\title{Quantitative constraint-based computational model of tumor-to-stroma coupling via lactate shuttle}

\author{Fabrizio Capuani}
\affiliation{Dipartimento di Fisica, Sapienza Universit\`a di Roma, Rome (Italy)}

\author{Daniele De Martino}
\affiliation{Dipartimento di Fisica, Sapienza Universit\`a di Roma, Rome (Italy)}
\affiliation{Center for Life Nano Science@Sapienza, Istituto Italiano di Tecnologia, Rome (Italy)}

\author{Enzo Marinari}
\thanks{Authors contributed equally}
\affiliation{Dipartimento di Fisica, Sapienza Universit\`a di Roma, Rome (Italy)}
\affiliation{INFN Sezione di Roma 1, Rome (Italy) }

\author{Andrea De Martino}
\thanks{Authors contributed equally}
\affiliation{Dipartimento di Fisica, Sapienza Universit\`a di Roma, Rome (Italy)}
\affiliation{Center for Life Nano Science@Sapienza, Istituto Italiano di Tecnologia, Rome (Italy)}
\affiliation{Soft and Living Matter Laboratory, Istituto di Nanotecnologia (CNR/NANOTEC), Consiglio Nazionale delle Ricerche, Rome (Italy)}

\begin{abstract}
Cancer cells utilize large amounts of ATP to sustain growth, relying primarily on non-oxidative, fermentative pathways  for its production. In many types of cancers this leads, even in the presence of oxygen, to the secretion of carbon equivalents (usually in the form of lactate) in the cell's surroundings, a feature known as the Warburg effect. While the molecular basis of this phenomenon are still to be elucidated, it is clear that the spilling of energy resources contributes to creating a peculiar microenvironment for tumors, possibly characterized by a degree of toxicity. This suggests that mechanisms for recycling the fermentation products (e.g. a lactate shuttle) may be active, effectively inducing a mutually beneficial metabolic coupling between aberrant and non-aberrant cells. Here we analyze this scenario through a large-scale {\it in silico}  metabolic model of interacting human cells. By going beyond the cell-autonomous description, we show that elementary physico-chemical constraints indeed favor the establishment of such a coupling under very broad conditions. The characterization we obtained by tuning the aberrant cell's demand for ATP, amino-acids and fatty acids and/or the imbalance in nutrient partitioning provides quantitative support to the idea that synergistic multi-cell effects play a central role in cancer sustainment.
\end{abstract}

\maketitle

\section*{Introduction}

At heart, a cell's energetic problem consists in selecting how to process nutrients (say, glucose molecules) into chemical energy (adenosine 5'-triphosphate, ATP) that will then be transduced into useful forms of mechanical or chemical work. Rapid cellular growth, in specific, requires high rates of macromolecular biosynthesis and of energy production, which presupposes (a) fast ATP generation, and (b) tight control of the cell's redox state, i.e. that the ratio between the levels of electron donors and acceptors stays in a range that guarantees functionality. Most often, molecular oxygen is the primary electron acceptor in cells,  playing a central role in the electron transfer chain (ETC) that constitutes the main ATP-producing mechanism in cells.  When a glucose molecule enters the cell, it is normally metabolized by glycolysis, a highly conserved reaction pathway that converts each glucose anaerobically into two molecules of pyruvate, with the concomitant production of 2 ATPs. In presence of oxygen, cells can operate the ETC, which begins with the conversion of pyruvate into acetyl-coenzyme-A (acetyl-CoA). The reaction pathways responsible for the subsequent production of ATP (and of many macromolecular precursors like amino-acids) are the Tricarboxylic Acid (TCA) cycle and Oxidative Phosphorylation (OXPHOS). These complex groups of reactions (roughly 100 processes altogether in the bacterium E. coli) are able to generate the largest energy yield in terms of molecules of ATP produced per glucose molecule intaken (up to 36, adding to the 2 given by glycolysis), and release carbon dioxide as a waste product. In absence of oxygen, however, cells cannot rely on the ETC and the ATP yield of glycolysis (2) is to a good approximation all the energy they can generate. In such conditions, the pyruvate obtained from glycolysis is then reduced to other carbon compounds  (e.g. acetate, ethanol, lactate) that are normally excreted in variable amounts. The conversion of pyruvate to lactate, is carried out by a single reaction catalyzed by the enzyme lactate dehydrogenase (LDH). 

The energy-generating strategies just described are, in a sense, the two extremes, and cells usually operate mixtures of the two even in presence of oxygen, leading to ATP yields below the theoretical maximum of 38 (typically around 30). However, fast proliferating cells normally display high rates of glucose intake and produce ATP anaerobically even in the presence of oxygen, thereby spilling potentially useful carbon and energy resources.
A hint about why a large glucose influx may favor the use of lower-yield pathways  is provided by the fact that processing high glucose fluxes via glycolysis requires high rates of production of adenosine 5'-diphosphate (ADP) and of NAD$^+$, via oxidation of NADH. The simplest way to convert NADH back into NAD$^+$ is by reduction of pyruvate to lactate via LDH. Therefore sustaining high rates of glucose metabolization may imply lactate overflow. This however seems to suggest that a cell with a large glucose intake should always prefer to generate energy by glycolysis. Therefore, different constraints (physical, regulatory, thermodynamic, etc.) may be at work in the selection of a cell's energetic strategy \cite{Molenaar:2009p3978}. We note that recent high-throughput studies of the compounds secreted by growing bacteria in controlled environments (the so-called exo-metabolome) uncovered that, besides the standard outputs of overflow metabolism, a previously unsuspected diversity of molecules accompanies the excretion of carbon equivalents \cite{Paczia:2012p4266}.

Likewise, aerobic glycolysis with lactate overflow (a.k.a. Warburg effect) is found to occur in many types of cancers \cite{Hsu:2008p4267, Kroemer:2008p3885}, although it cannot be considered as a necessary signature of malignancy \cite{Funes:2007p4268}. In order to explain its predominance at the molecular level, several ideas have been pursued, from structural or genetic abnormalities in the mitochondria \cite{Zhou:2007p4269}, to the roles played by the hypoxia-inducible factor (HIF) \cite{Pouyssegur:2006p4277} and by a specific isoform of the glycolytic enzyme pyruvate kinase \cite{Christofk:2008p3930}, to seemingly unrelated genetic events leading to the increase of glucose transporters and the loss of growth control in cancer cells \cite{Levine:2010p4275}. The search for a molecular basis of the Warburg effect has given momentum to the idea that treatments targeting the simpler energy-producing apparatus of a cell (as opposed to its complex and not entirely known genetic profile) carry a higher potential than gene-based therapies \cite{Tennant:2010p3811}. Unluckily, our understanding of cancer metabolism is still largely incomplete \cite{VanderHeiden:2009p3807, Cairns:2011p3989}, and research on the role of metabolism in tumor genesis and progression is currently undergoing a major revival \cite{Dang:2012p4276}.

Cell-autonomous models of cellular metabolism based on genome-scale reconstructions of the underlying  network of metabolic reactions have been widely studied, proving effective in explaining the origin of overflow metabolism in different contexts. Constraint-based approaches like Flux-Balance Analysis (FBA), for instance, have shed light on the roles that different types of constraints may play in driving the selection of energetic strategies towards aerobic glycolysis. An interesting and intuitively appealing suggestion is put forward by the macromolecular crowding scenario \cite{Vazquez:2010p3868}, according to which enzymes carrying out energy-efficient pathways can at most occupy a fraction of the intracellular volume (e.g. the mitochondria in eukaryotes). Because reactions are assumed to be enzyme-limited, the activity of aerobic pathways will necessarily level out once all deputed enzymes work at full speed, while that of glycolysis may still increase. These models are indeed able to predict carbon excretion by unicellulars \cite{Famili:2003p4281} as well as the Warburg effect in cancer cells \cite{Shlomi:2011p3812, Vazquez:2010p3868}. Many issues however deserve further analysis.

One in particular concerns the fact that cellular waste products such as lactate tend to deteriorate the extracellular environment, so that the sustainability of aerobic glycolysis depends crucially on the possibility that such a pollution is remediated within cancer's microenvironment. 
This raises questions about the role of tumor-stroma \rev{metabolic} interactions. Tumors are universally characterized by a marked upregulation of glucose transporters \cite{Levine:2010p4275, Locasale:2009p3927}, which allows them to largely surpass their non-aberrant (stromal) neighbors in the competition for nutrients. As a consequence, stromal cells must adjust their energetics in ways that have only started to be analyzed \cite{Deberardinis:2012p4007}. 

Lactate exchanges are increasingly being recognized to play an important role in this respect. In a scenario that has been characterized by genetic signatures over the past few years, tumor-derived lactate can be taken up by glucose-starved cancer-associated fibroblasts (CAFs) \cite{pmid16423989}, which would then both help their own survival and foster cancer's aberrant growth by sanitizing the environment. Here  we present a quantitative {\it in silico} analysis of the above mechanism. Our results confirm that tumor-to-stroma lactate shuttling appears as a robust consequence of basic physical and chemical constraints, in a manner that is largely independent of (i) the metabolic function (e.g. energy production, amino-acid synthesis, fatty acid synthesis, etc.) that aberrant cells strive to maximize and, perhaps more surprisingly, (ii) of the degree to which they manage to maximize it. The core of our conclusions can already be reached through a highly simplified model of metabolism that captures the bare essential features of the mechanism. The intuition is then validated on  a large-scale model of catabolism based on  recent reconstructions of the human reactome. After characterizing the network at the cell-autonomous level, we consider a system formed by two such networks sharing the same glucose supply, focusing on the case in which one cell aims to maximize a growth-related objective function (e.g. energy production) with the other subject to a simple ATP maintenance constraint. The emerging scenario is studied by tuning the `degree of maximization' by the aberrant cell through a rigorous sampling scheme that provides a statistically significant description of all feasible metabolic phenotypes compatible with the constraints. Lactate overflow and functional tumor-stroma coupling appears robustly upon maximizing production of ATP as well as of several biomass precursors.

It is important to stress that a tumor-to-stroma coupling, while compatible with the observation that the vast majority of cancer cells display glycolytic metabolism, is not the only lactate-based shuttling postulated to be relevant for tumor progression (see e.g. \cite{Pavlides:2009p3926}). Other mechanisms and their relation to the current study are outlined in the Discussion.

The remainder of the manuscript is organized as follows. The Results section begins with a brief summary of a stylized single-cell model of aerobic glycolysis presented in \cite{Vazquez:2010p3868}, where it is shown how a crowding constraint can divert ATP production from oxidative phosphorylation to fermentation. The simple model is then extended to two cells whereby the extent of lactate shuttle is computed analytically. A large scale metabolic network of the human core metabolism (HCCN) is then introduced. A rigorous sampling of the metabolisms produced by HCCN confirms the possibility of cell coupling by means of lactate shuttle between a cancer cell and a stromal cell.
The rigorous samplings allows us to compute the correlation coefficients among the fluxes of the cancer and stromal cell. In the Discussion section, experimental evidences for lactate shuttle are presented and, finally, physiological implications and possible experimental validation of the results are discussed.

\section*{Results}

\rev{\subsection*{Metabolic fluxes and crowding constraint in a minimal cell-autonomous model}}

Our starting point is the highly simplified cell-autonomous model defined in \cite{Vazquez:2010p3868}, in which lactate secretion is related to an {\it ad hoc} constraint limiting volume available for energy production in cells that maximize the ATP output flux. In particular, when ATP demands exceed the volume allocated for ATP production, the cell produces ATP also through fermentation. In the model of Vazquez et al., a cell is characterized by a glucose intake flux $U_{\G,{\rm in}}$,  a glycolytic flux  $f_{\GL}$ (yielding two pyruvate molecules per intaken glucose) and an overall mitochondrial OXPHOS flux $f_{\M}$ that transforms pyruvate into energy (a table summarizing the most relevant variables defined here is given as Table S1). Alternately, pyruvate can be turned into lactate by lactate dehydrogenase (LDH), with flux $f_{\L}$. Lactate can then be expelled from the cell so as to avoid acidification. Mass balance and steady state conditions impose that metabolite levels do not change in time, so that, for example, the net production and consumption fluxes of pyruvate must compensate, implying
\begin{equation}\label{unoo}
2f_{\GL}-f_{\M}-f_{\L}=0~~,
\end{equation}
(note that, to make the  subsequent analysis simpler, our algebra differs slightly in numerical pre-factors from the one given in \cite{Vazquez:2010p3868}. The overall picture is however completely equivalent).
The glycolytic flux is in turn limited by the flux $U_{\G}$ that supplies glucose to the system, so that $U_{\G,{\rm in}}\leq U_{\G}$, while under mass balance $U_{\G,{\rm in}}=f_{\GL}$.
The crowding constraint is implemented as $V_{\GL}+V_{\M}+V_{\L}\leq V_{\ATP}$, where $V_{\ATP}$ represents the cell volume devoted to ATP production, whereas $V_{\GL}$, $V_{\M}$, and $V_{\L}$ are the volumes taken up by the glycolytic, OXPHOS, and LDH enzymes, respectively. Introducing the constants $a_{\GL},$ $a_{\M},$ and $a_{\L}$ representing the volume occupied per unit of ATP production by glycolytic, OXPHOS and LDH enzymes, respectively, this can be re-cast as \cite{Vazquez:2010p3868}
\begin{equation}
a_{\GL}f_{\GL}+a_{\M}f_{\M}+a_{\L}f_{\L}\leq\Phi_{\ATP}~~,
\label{ORIGINALVOLUMECONSTRAINT}
\end{equation}
where $\Phi_{\ATP}$ is the volume fraction available for ATP production in each cell. 
Using (\ref{unoo}), one obtains
\begin{equation}
\left(a_{\L}+\frac{a_{\GL}}{2}\right)f_{\L}+\left(a_{\M}+\frac{a_{\GL}}{2}\right)f_{\M}\leq\Phi_{\ATP}~~.\label{volumeoccupied}
\end{equation}
We shall adhere to \cite{Vazquez:2010p3868} in employing the empirical estimates  $a_{\GL}\simeq 3\times10^{-3}$ (min/mM), $a_{\L}\simeq 4.6\times10^{-4}$ (min/mM), $a_{\M}\simeq 2\times10^{-1}$ (min/mM), and $\Phi_{\ATP}=0.4$.
\rev{Equation \eqref{volumeoccupied} represents the ``crowding constraint'', amounting to a global constraint to the overall flux of metabolites the cell can invest in ATP production. To be consistent with \cite{Vazquez:2010p3868}, we derived the crowding constraint on fluxes by limiting the volume fraction devoted to energy production, but the constraint \eqref{volumeoccupied} might as well be derived by limiting the amount of proteins (see, e.g., \cite{Shlomi:2011p3812}).}

\rev{To assess the amount of ATP produced by these fluxes, we consider}
 that $f_{\GL}$ generates 2 ATPs and 2 pyruvates per glucose molecule, and that $f_{\M}$ creates 18 ATP molecules from each pyruvate. \rev{The conversion of pyruvate to lactate does not generate any ATP and it only ensures that glycolysis can continue by regenerating NADH to NAD.} The overall flux of ATP  production is finally given by 
\begin{equation}
f_{\ATP}=2f_{\GL}+18f_{\M}~~.\label{eq: ATPflux}
\end{equation}

Vazquez et al's results can be summarized as follows. Neglecting \eqref{volumeoccupied}, ATP production is maximized when $f_{\GL}=U_{\G}$, $f_{\M}=2f_{\GL}=2U_{\G},$ and $f_{\L}=0$, i.e. when all available glucose is taken up by the cell and used in OXPHOS. This solution still holds if  \eqref{volumeoccupied} is satisfied, i.e. as long as
\begin{equation}
\left(2a_{\M}+a_{\GL}\right)U_{\G}\le\Phi_{\ATP}~~.
\end{equation}
If however $U_{\G}>\Phi_{\ATP}/\left(2a_{\M}+a_{\GL}\right)\equiv u_{\G}$, then the cellular \rev{resources} available for high-yield energy production pathways has been exhausted and ATP synthesis \rev{is no longer limited by glucose but by intracellular resources (either volume or proteins production capacity).}
Its maximization now requires that the glucose flux exceeding the possibility of processing by OXPHOS is diverted towards LDH. The maximum ATP production flux and the corresponding lactate flux for this case can be obtained by inserting $f_\L=2U_{\G}-f_\M$ into \eqref{volumeoccupied} with the constraint saturated. One gets
\begin{gather}\label{lac}
f_{\L}=
\begin{cases}
0 &\text{if $U_\G \le u_{\G}$}\\
\alpha_\M (U_{\G}-u_{\G}) &\text{if $U_\G>u_{\G}$}
\end{cases}~~,
\\
\alpha_\M= \frac{2a_{\M}+a_{\GL}}{a_{\M}-a_{\L}}>0\nonumber
\end{gather}
and
\begin{gather}
f_{\M}=
\begin{cases}
2U_\G &\text{if $U_\G \le u_{\G}$}\\
\alpha_\L(v_{\G}-U_{\G}) &\text{if $U_\G>u_{\G}$}
\end{cases}~~,
\\
\alpha_\L= \frac{2a_{\L}+a_{\GL}}{a_{\M}-a_{\L}}>0\nonumber
\label{eq:fM}
\end{gather}
where 
$v_{\G}=\Phi_{\ATP}/\left(2a_{\L}+a_{\GL}\right)$.
Because of mass balance, $f_\L$ also equals the lactate excretion flux so that the Warburg effect is described by (\ref{lac}), which shows the existence of a threshold glucose intake above which lactate is secreted.

\rev{\subsection*{Metabolic fluxes in a minimal model of two cells coupled via a lactate shuttle}}

\begin{figure}
\centering
\includegraphics[width=0.5\textwidth]{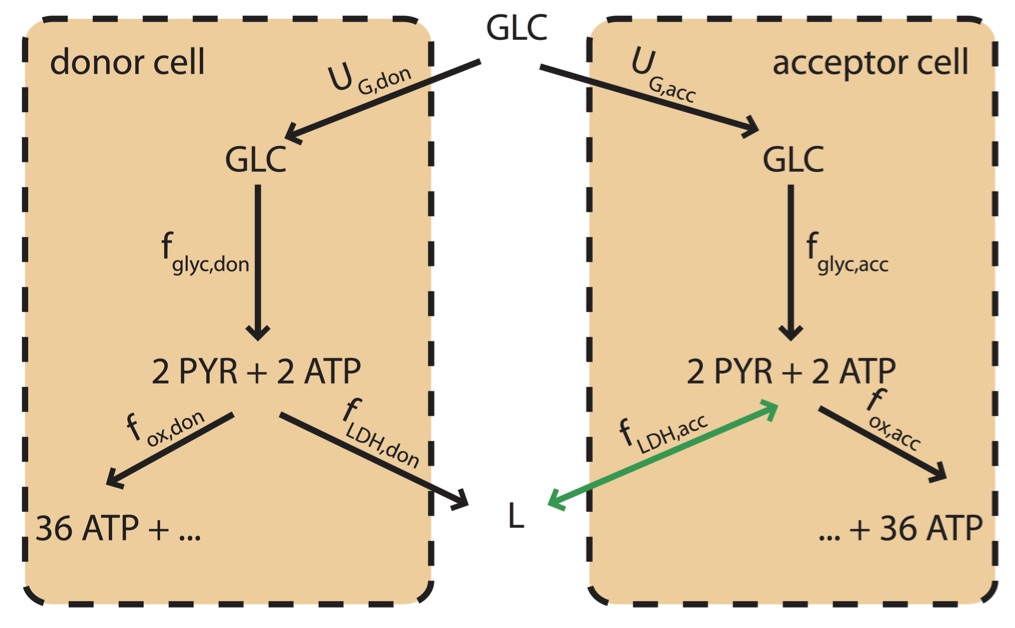}
\caption{\textbf{Schematic representation of the minimal model of two cells coupled via a lactate shuttle.} Two cells (lactate donor and acceptor, respectively) share glucose as an energy source. Glucose is partitioned according to the fluxes $f_{\GL,\D}$ and $f_{\GL,\A}$, which convert one internal glucose molecules to two pyruvate molecules  producing two ATP molecules. Pyruvate can then undergo oxidative phosphorylation (with the irreversible fluxes $f_{\M,\D}$  and $f_{\M,\A}$, giving 36 more ATP molecules) or LDH (with fluxes $f_{\L,\D}$ and $f_{\L,\A}$, by which lactate (L) is obtained). Lactate is for simplicity assumed to be secreted upon production. If both cells produce lactate, no coupling sets in, unless due to competition for nutrients under glucose limitation. If however the donor cell secretes lactate, the acceptor cell may intake it to replace glucose whenever its access to the latter is limited (e.g. because the donor cell's glucose intake is large). In such cases a lactate shuttle will effectively couple the metabolisms of the two cells.}
\label{FIG: two Vazquez-like cells}
\end{figure}

Let us now consider two replicas of Vazquez et al.'s cell and assume that the two cells can interact via the exchange of lactate, which, in particular, can be intaken and used as an alternative carbon source through reverse LDH (catalyzing the conversion of lactate back to pyruvate). For sakes of simplicity, we distinguish between a donor cell (`don', which can only excrete lactate) and an acceptor cell (`acc', which can both excrete and import lactate) (see Fig.~\ref{FIG: two Vazquez-like cells}).
In order to avoid starvation, we furthermore impose that both cells generate a minimum flux $\atpmin$ of  ATP, i.e.
\begin{equation}
f_{\ATP,i}\geq \atpmin ~~\label{eq: no Starvation},~~~~~ i\in\{\A, \D\}
\end{equation}
where the ATP production flux is given by (\ref{eq: ATPflux}), namely 
\begin{equation}
f_{\ATP,i}=2f_{\GL,i}+18 f_{\M,i}~~~~~  i\in\{\A, \D\}~~.
\label{eq:ATPproduction2cells}
\end{equation}
\rev{As shown in the beginning of the previous section, when $f_{\GL}=U_{\G}$ and $f_{\M}=2U_{\G}$ cells can extract the maximum ATP from a single glucose molecule, i.e., $f_{\ATP,i}=38U_{\G}$.}
We observe that, in order to satisfy \eqref{eq: no Starvation}, the overall glucose supply to the system \rev{composed of two cells} should satisfy \begin{equation}
U_{\G}\geq \frac{ 2 \atpmin}{38}\equiv 2 u_{\G,0}~~,\label{eq: Minimal Glucose Flux}
\end{equation}
which defines $u_{\G,0}$ as the minimum glucose supply flux that can prevent cellular starvation.
Denoting as $U_{\G,\D}$  the amount of glucose available to the donor cell, if the donor cell maximizes ATP production, its flux organization will match the one found above for a single cell.
It is useful to distinguish two limiting cases:
\begin{itemize}
\item[(a)] If the donor cell employs  OXPHOS exclusively, which happens when $2u_{\G,0}\leq U_\G\leq u_{\G}+u_{\G,0}$,
we have $U_{\G,\D}=U_{\G}-u_{\G,0}$ and $U_{\G,\A}=u_{\G,0}$. In other words, the donor cell takes up all of the available glucose except for the amount necessary for the survival of the acceptor cell. In these conditions, the acceptor cell's survival is guaranteed by glucose availability. 
\item[(b)] At the other extreme, if the acceptor cell can survive  even using exclusively the lactate excreted by the donor cell, all glucose is intaken by the donor cell, so that $U_{\G,\D}=U_\G$ and $U_{\G,\A}=0$. This happens when $f_{\L,\D}\ge \atpmin/18$, or 
\begin{equation}
U_\G\ge \frac{\atpmin}{18 \alpha_\M}+u_\G~~.
\end{equation}
\end{itemize}
In intermediate situations, the amount of glucose intaken by the acceptor cell gradually decreases from $u_{\G,0}$ to 0 as it increasingly uses lactate to produce pyruvate and thus ATP. In such cases, the oxidative pathway of the acceptor cell is fed by both glucose and lactate fluxes as $f_{\M,\A}=2f_{\GL,\A}+f_{\L,\D}$, which, via mass balance, implies (see \eqref{eq:ATPproduction2cells})
\begin{equation}
\label{minATP_intermediate}
f_{\ATP,\A}=38f_{\GL,\A}+18 f_{\L,\D}=38 U_{\G,\A} + 18 f_{\L,\D}~~.
\end{equation}
To obtain simpler expressions, it is useful to introduce the shorthands
\begin{gather}
u_{\G,1}=u_{\G,0}+u_\G ~~, \\
u_{\G,2}=\frac{\atpmin}{18 \alpha_\M}+u_\G~~,
\end{gather}
and $\Delta U_{\G}$ for the amount of glucose rerouted from the acceptor cell to the donor cell, so that
\begin{equation}
\label{intermediate_glucose_fluxes_implicit}
\begin{aligned}
U_{\G,\A} &=u_{\G,0}-\Delta U_{\G}\\
U_{\G,\D} &=U_{\G}-u_{\G,0}+\Delta U_{\G}~~.
\end{aligned}
\end{equation}
Substituting $U_{\G,\A}$ and imposing minimal ATP requirement in \eqref{minATP_intermediate}, we obtain
\begin{equation}
\label{lac_secretion_implicit}
f_{\L,\D}=\frac{38}{18}\Delta U_{\G}~~.
\end{equation}
Since the donor cell maximizes ATP production, we obtain a second expression for $f_{\L,\D}$ as a function of $\Delta U_{\G}$ by substituting $U_{\G,\D}$:
in \eqref{lac}
\begin{equation*}
f_{\L,\D}=\alpha_\M ( U_{\G,\D} - u_{\G}) = \alpha_\M ( U_\G-u_{\G,0}+\Delta U_{\G} - u_{\G})~~.
\end{equation*}
This, combined with \eqref{lac_secretion_implicit}, determines the extra amount of glucose available to the donor cell
\begin{equation}
\label{extra_glucose}
\Delta U_{\G} = \frac{\alpha_\M}{38/18-\alpha_\M}\left[ U_{\G}-u_{\G,0}-u_{\G}\right]~~.
\end{equation}
Expression \eqref{extra_glucose} is valid if $u_{\G,1} < U_\G\leq u_{\G,2}$. For $U_\G \le u_{\G,1}$ one has $\Delta U_\G=0$, while when $U_\G> u_{\G,2}$ $\Delta U_\G=u_{\G,0}$.
We can therefore summarize the fluxes for the donor cell as
\begin{gather}\label{lac2}
f_{\L,\D}=
\begin{cases}
0 &\text{if $2u_{\G,0}\leq U_\G\leq u_{\G,1}$}\\
\alpha_\M (U_{\G}+\Delta U_{\G}-u_{\G,1}) &\text{if  $u_{\G,1} < U_\G\leq u_{\G,2}$}\\
\alpha_\M (U_{\G}-u_{\G}) &\text{if  $U_\G > u_{\G,2}$}
\end{cases}
\end{gather}
and
\begin{gather}
f_{\M,\D}=
\begin{cases}
2(U_\G-u_{\G,0}) &\text{if $2u_{\G,0}\leq U_\G \leq u_{\G,1}$}\\
\alpha_\L(v_{\G}-U_{\G,\D}) &\text{if  $u_{\G,1} < U_\G\leq u_{\G,2}$}\\
\alpha_\L(v_{\G}-U_{\G}) &\text{if  $U_\G > u_{\G,2}$}
\end{cases}~~.
\end{gather}
Because the donor cell uses up most of the glucose, the acceptor cell stays at the minimum ATP production rate $\atpmin$ for $U_\G\leq u_{\G,2}$. In this regime, the fluxes of the acceptor cell are fixed and can be derived from the ones of the donor cell. One gets
\begin{gather}\label{lac2_cell2}
f_{\L,\A}=
\begin{cases}
0 &\text{if $2u_{\G,0}\leq U_\G\leq u_{\G,1}$}\\
-f_{\L,\D} &\text{if  $u_{\G,1} < U_\G\leq u_{\G,2}$}
\end{cases}
\end{gather}
and
\begin{gather}
f_{\M,\A}=
\begin{cases}
2u_{\G,0} &\text{if $2u_{\G,0}\leq U_\G \leq u_{\G,1}$}\\
2U_{\G,\A}+f_{\L,\D} &\text{if  $u_{\G,1} < U_\G\leq u_{\G,2}$}
\end{cases}~~.
\end{gather}
For large total glucose intakes, the donor cell excretes more lactate than necessary for bare survival of the acceptor cell, which can produce any amount of ATP as long as it is larger than $\atpmin$. The metabolic state of the acceptor cell is therefore not uniquely defined. When $U_\G\ge u_{\G,2}$, for the fluxes of the acceptor cell one finds
\begin{gather}\label{lac2_partial_intake}
-f_{\L,\D} \leq f_{\L,\A}\leq -\frac{\atpmin}{38} \\
f_{\M,\A} = -f_{\L,\A}~~,
\end{gather}
while the ATP production is given by
\begin{equation}
f_{\ATP,\A}  =  -18f_{\L,\A}~~.
\end{equation}
Note that $f_{\L,\A}<0$ because the acceptor cell intakes lactate and reverses LDH. 

\begin{figure}
\centering
\includegraphics[width=0.5\textwidth]{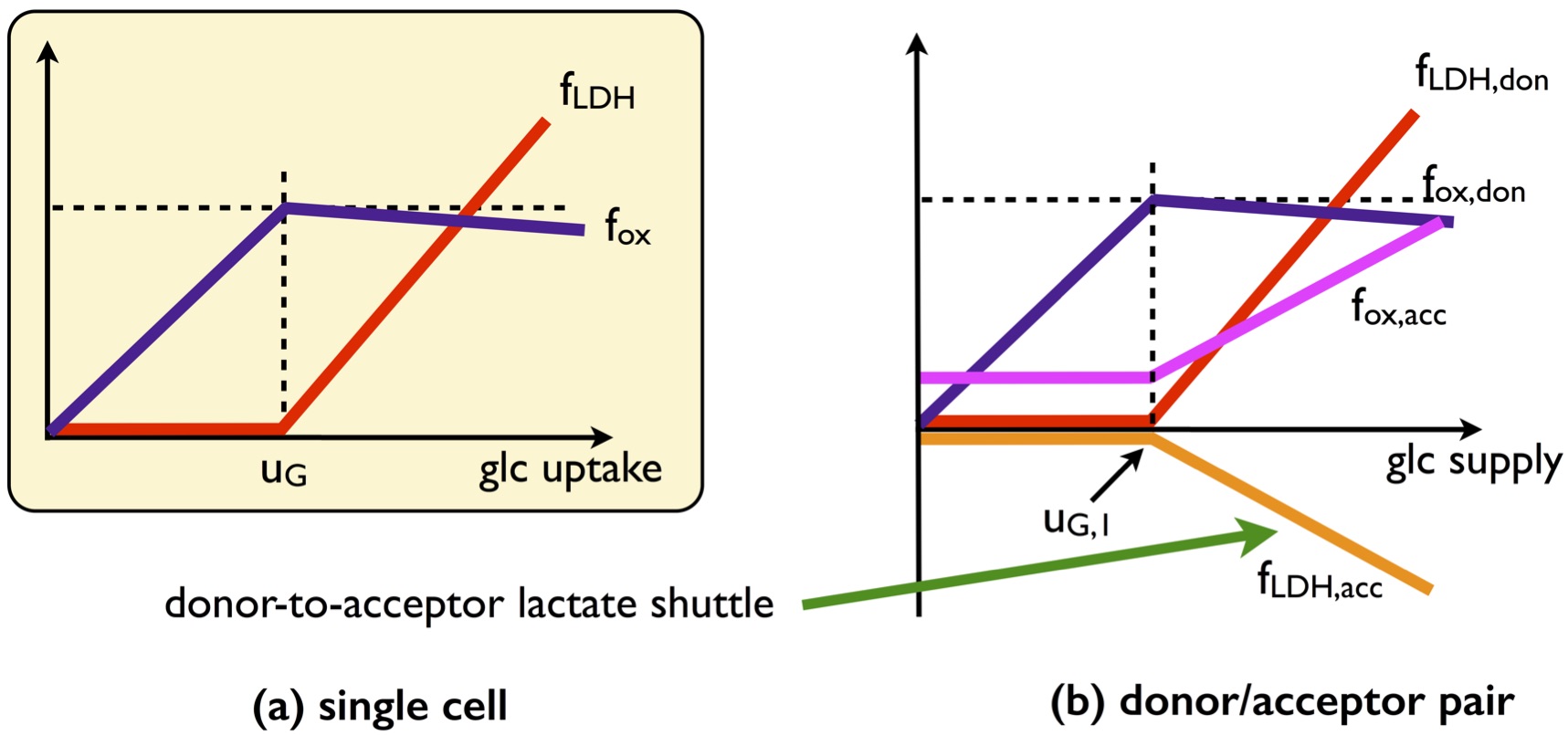}
\caption{\textbf{Qualitative behaviour of the single-cell and of the donor-acceptor system for the coarse-grained model.} (a) In the cell-autonomous model, lactate overflow occurs when the glucose intake overcomes a threshold. Correspondingly, the flux through oxidative metabolism increases until the threshold before slowly decreasing once the \rev{crowding} constraint is saturated and fermentation sets in. (b) In the donor/acceptor system, the donor behaves essentially as an autonomous cell and the acceptor adapts to it. For low glucose intakes, it operates oxidative pathways at small rate. When the donor saturates the \rev{crowding} constraint excreting lactate, the acceptor imports it and uses it at a substrate to increase the flux through oxidative metabolism.}
\label{picture}
\end{figure}

\begin{figure*}
\centering
\includegraphics*[width=0.75\textwidth]{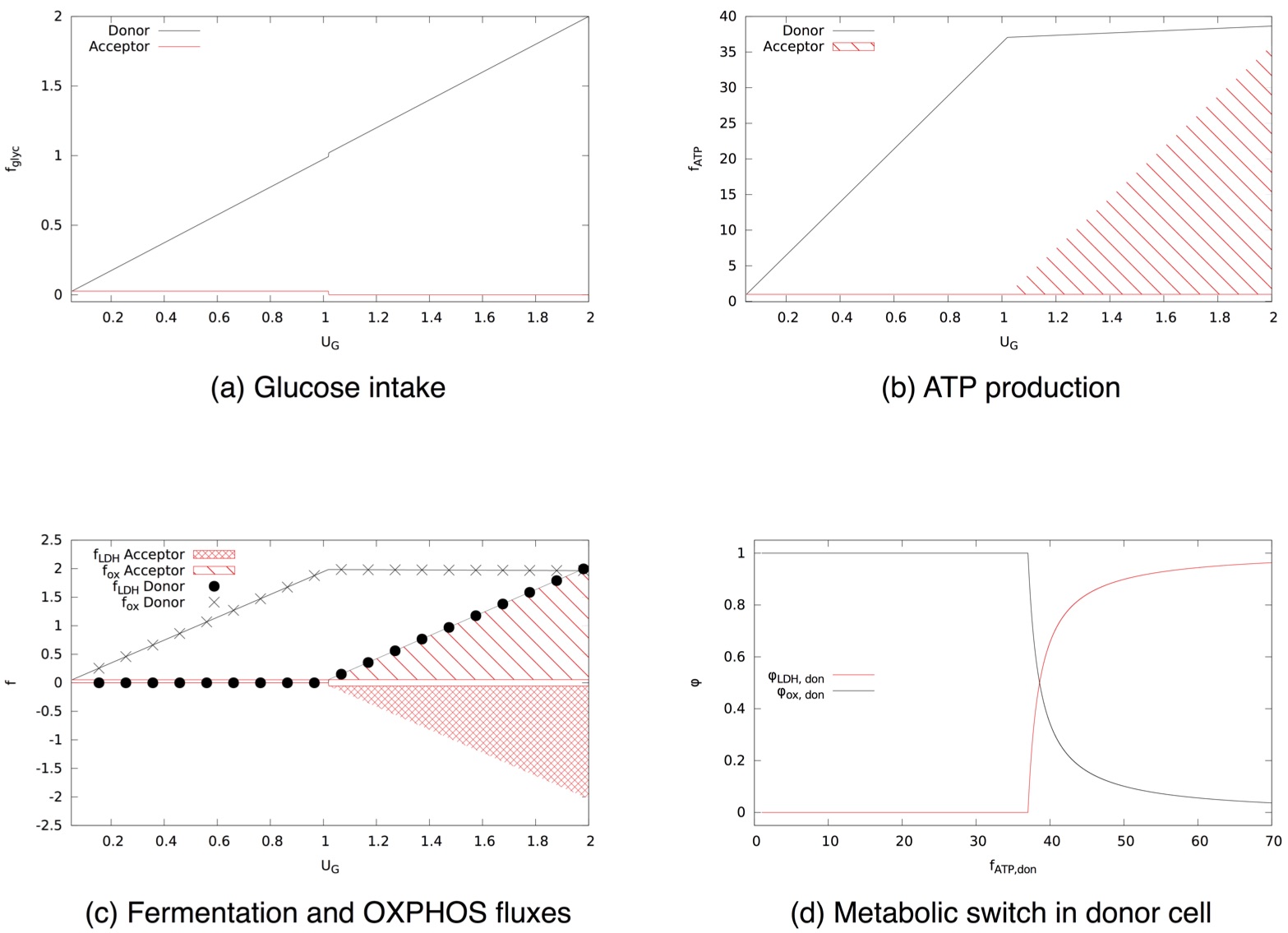}
\caption{\textbf{Solution of the minimal model of two cells coupled via a lactate shuttle.}
A lactate donor cell that maximizes ATP production is coupled to an acceptor cell.
(a) Glucose intake for donor and acceptor as a function of total glucose supply. 
(b) ATP production as a function of the total glucose supply. The shaded area indicates that all values within that region are feasible.
(c) Flux through fermentation $f_{\L}$ (circles) and oxidative phosphorylation $f_{\M}$ (crosses) as a function of the total glucose supply.
(d) Fraction of ATP produced via fermentation (black) or via oxidative phosphorylation (red) in the donor cell as a function of the total ATP  it produces.}
\label{fig: analytic sampling of LA cell}
\end{figure*}

The solution to this model is described pictorially in Fig.~\ref{picture} and discussed in Fig.~\ref{fig: analytic sampling of LA cell}, where we also show the feasible range of values of the fluxes of the acceptor cell. 

In essence, this coarse-grained model suggests that an imbalance in the energetic demands of two cells can induce a metabolic coupling driven by a lactate shuttle from the high- to the low-demand cell, and provides a qualitative scenario for how carbon utilization in both cells changes by tuning the overall glucose supply. We shall now see that such a picture is fully recovered within a large-scale model of human metabolism.

\rev{\subsection*{Metabolic fluxes and crowding constraint in a cell-autonomous large scale metabolic network}}


In order to characterize the exchange of carbon equivalents among interacting cells more in detail, we have analyzed a human metabolic network derived from the reconstructed human reactome \cite{Duarte:2007p3938}, to which we refer as the `Human Core Catabolic Network' (HCCN, see `Materials and Methods'). The HCCN is composed by 67 metabolites and 75 reactions (including uptake fluxes), 23 of which are reversible.  The \rev{crowding} constraint that accounts for finite cellular resources is represented as
\begin{equation}
a_{\GL}f_{\HEX}+a_{\M}(f_{\PDH}+{ f_{\GLUN}})+a_{\L}f_{\LDH}\leq\Phi_{\ATP}\label{eq: HCCN volume constraint}
\end{equation}
where $f_{\HEX}$ denotes the Hexokinase-1 flux, which irreversibly channels glucose into glycolysis, $f_{\PDH}$ denotes the flux through pyruvate dehydrogenase, by which pyruvate is diverted into the TCA cycle  and $f_{\GLUN}$ denotes the mithocondrial flux of nitrogen metabolism through glutaminase. We explore the space of flux patterns consistent with the constraints by sampling solutions of the mass balance equations $\mathbf{Sf=0}$ ($\mathbf{S}$ denoting the stoichiometric matrix and $\mathbf{f}$ being a flux vector) such that flux vectors $\mathbf{f}_\D$ of the donor cell appear with probability
\begin{equation}\label{gibbs}
P(\mathbf{f}_\D)= A~\exp[\beta f_{\ATP,\D}]~~,
\end{equation}
where $\beta\geq 0$ is a parameter and $A$ is a normalization constant. The rationale is that, by tuning the value of $\beta$, we can pass from an unbiased sampling in which $f_{\ATP,\D}$ takes on every allowed value with uniform probabilty ($\beta=0$) to one in which the donor cell maximizes its ATP output ($\beta\to\infty$), thereby obtaining a complete and refined picture of how different degrees of optimization by the donor cell impact the flux pattern of the acceptor cell.
\rev{Solutions at the two extremes can be obtained with standard methods, as linear programming, but only our method allows us to sample realistic sub-optimal optimizations. In Fig.~S6 we show how the ATP production increases as a function of $\beta$. In the following, as a prototype for sub-optimal optimization, we choose $\beta=5$ that corresponds to roughly $70\%$ of the maximum ATP production.}
Model and algorithmic details are discussed in `Materials and Methods'.

\begin{figure}
\centering
\includegraphics*[width=0.4\textwidth]{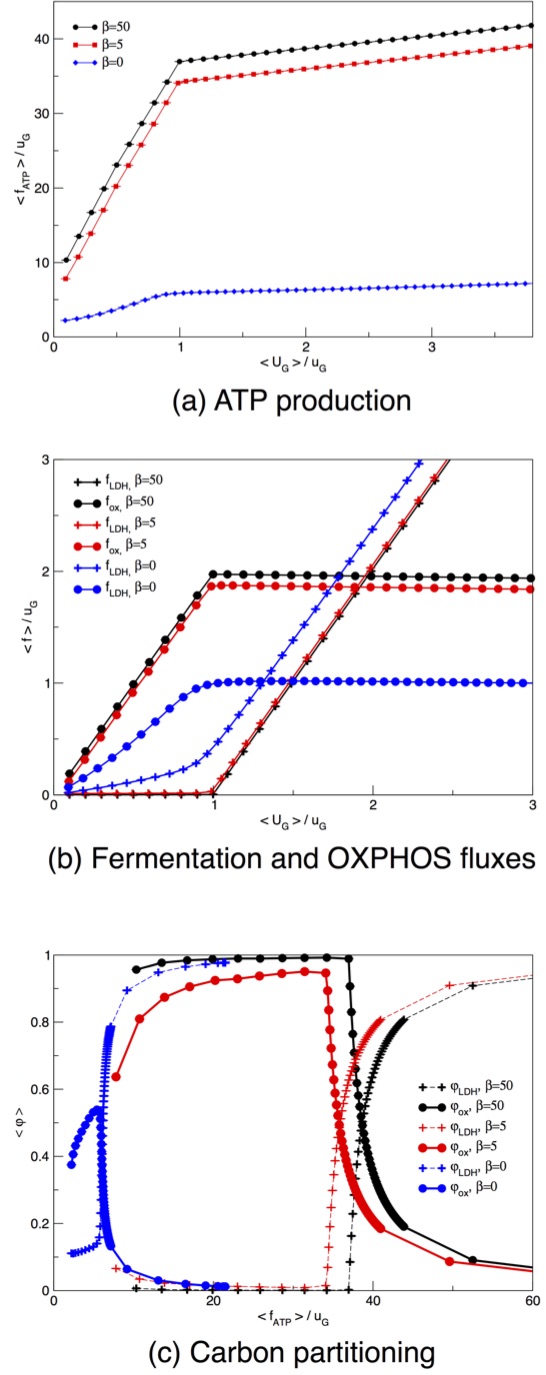}
\caption{\textbf{An isolated HCCN maximizes the ATP yield by directing glucose to OXPHOS even in absence of an active ATP maximization.}
(a) Average ATP production as a function of the re-scaled average glucose supply.
(b) Average flux through LDH (crosses) and PDHm (circles) as a function of the re-scaled average glucose supply.
(c) Fraction of ATP produced via glycolysis (crosses) or via OXPHOS (circles) as a function of the re-scaled total ATP produced.
The flux through each pathway is re-scaled by half the amount of glucose intaken by the cell (because with one molecule of glucose cells produce two molecules of pyruvate). Curves describe the behaviour of an ATP-maximizing HCCN (black lines,  $\beta=50$), a loosely maximizing donor (red lines, $\beta=5$) or the result of a uniform sampling of the allowed flux states for a HCCN (blue lines, $\beta=0$).
Error bars for the s.e.m. are smaller than symbols.}
\label{FIG: HCCN single cell}
\end{figure}

In Fig.~\ref{FIG: HCCN single cell}a we show the ATP production of a single HCCN cell as a function of the  available glucose. 
If ATP production is maximized (large $\beta$), one first encounters a regime with a yield of roughly 30 ATPs per glucose molecule (to be compared with the theoretical value of 38) and then a regime where the yield is about 1.7 ATPs per glucose. ATP is produced by OXPHOS in the former case, and by fermentation in the latter (see also Fig.~\ref{FIG: HCCN single cell}b and Fig.~\ref{FIG: HCCN single cell}c). As expected, networks maximizing ATP shift their metabolic strategy at the glucose intake for which the \rev{resources} available for ATP production by oxidation is exhausted. Note that for $\beta=5$ one obtains an ATP efficiency very close to optimal. Quite importantly and surprisingly, however, for $\beta=0$ (i.e. when no ATP maximization is performed) the emerging picture is qualitatively preserved, albeit with lower ATP yields.
Indeed solutions sampled at $\beta=0$ appear to employ a mixture of OXPHOS and fermentation even at low glucose intakes. It is remarkable that the \rev{resources}-driven shift still occurs at $u_{\G}$, as in this case it corresponds to a largely suboptimal value of ATP production. In other words, the HCCN might devote cellular resources to increase ATP production, but to do so it must explicitly be pursuing ATP maximization. This implies that the crossover from a high- to a low-yield strategy is a robust, embedded property of the network (and of the constraints imposed) and not an exclusive characteristic of the extremal solution that maximizes the ATP production.

In Fig.~\ref{FIG: HCCN single cell}b, we focus on the fluxes through PDHm and LDH, that indicate whether pyruvate is directed towards OXPHOS or fermentation. We see that high ATP yields are obtained by using OXPHOS exclusively. The typical network state at $\beta=0$ is less efficient at all glucose intakes, as it always diverts a larger fraction of pyruvate to fermentation compared to the ATP-maximizing cell. Finally, in Fig.~\ref{FIG: HCCN single cell}c, we detail how glucose partitions between fermentation and OXPHOS. The network displays a reversal of glucose fate as the \rev{crowding} constraint is saturated independently of whether ATP production is being maximized or not. The latter case however turns out to be generically less efficient.

\rev{\subsection*{Metabolic fluxes in a large scale metabolic network of two cells coupled via a lactate shuttle}}

\begin{figure}
\centering
\includegraphics*[width=0.45\textwidth]{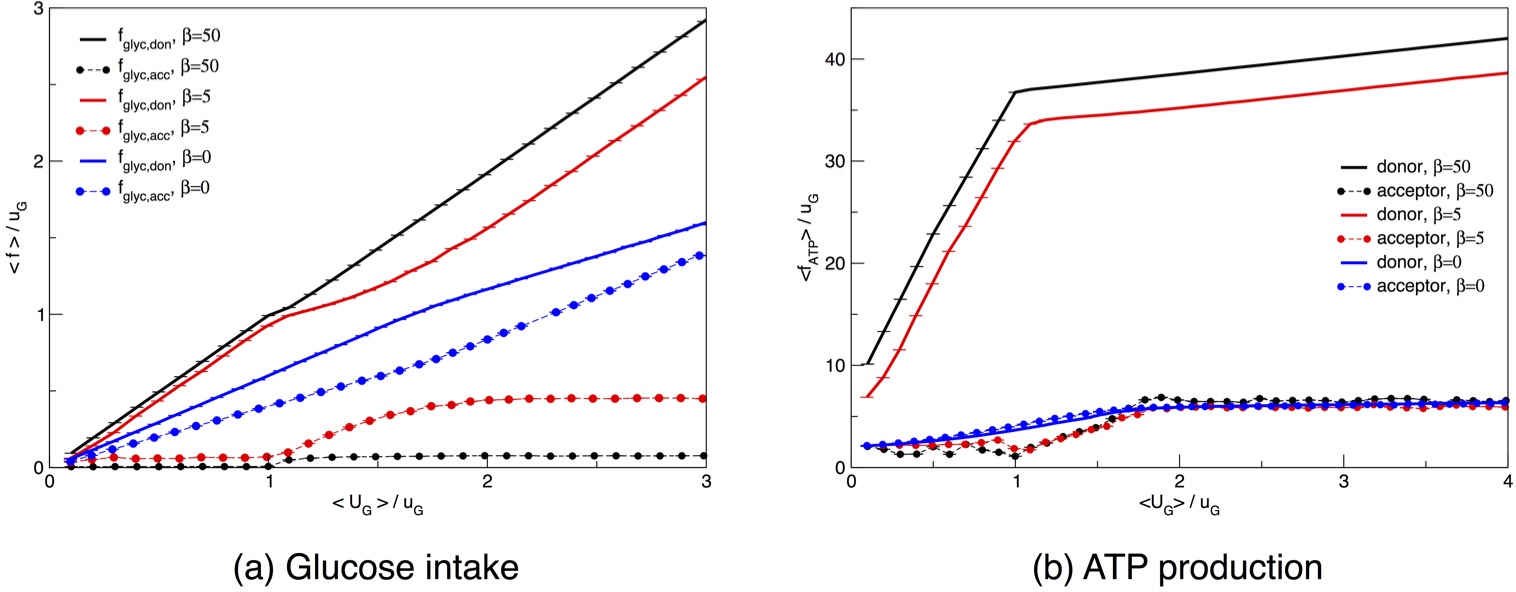}
\caption{
\textbf{When a lactate donor maximizes its ATP production it intakes most of the glucose supplied to the two-cell system. The ATP production by the acceptor cell increases in correspondence to a decrease in efficiency of the donor's metabolism.}
(a) Glucose intakes for two coupled cells as a function of the total glucose available to the donor-acceptor pair.
(b) ATP produced by the donor.
Curves describe the behaviour obtained for two coupled HCCN cells with an ATP-maximizing donor (black lines, $\beta=50$), a loosely maximizing donor (red lines, $\beta=5$) or for an unbiased sampling of the two-cell solution space (blue lines, $\beta=0$).
Error bars for the s.e.m. are smaller than symbols.}
\label{FIG: 2HCCN GLUC ATP}
\end{figure}

\begin{figure*}
\centering
\includegraphics*[width=0.7\textwidth]{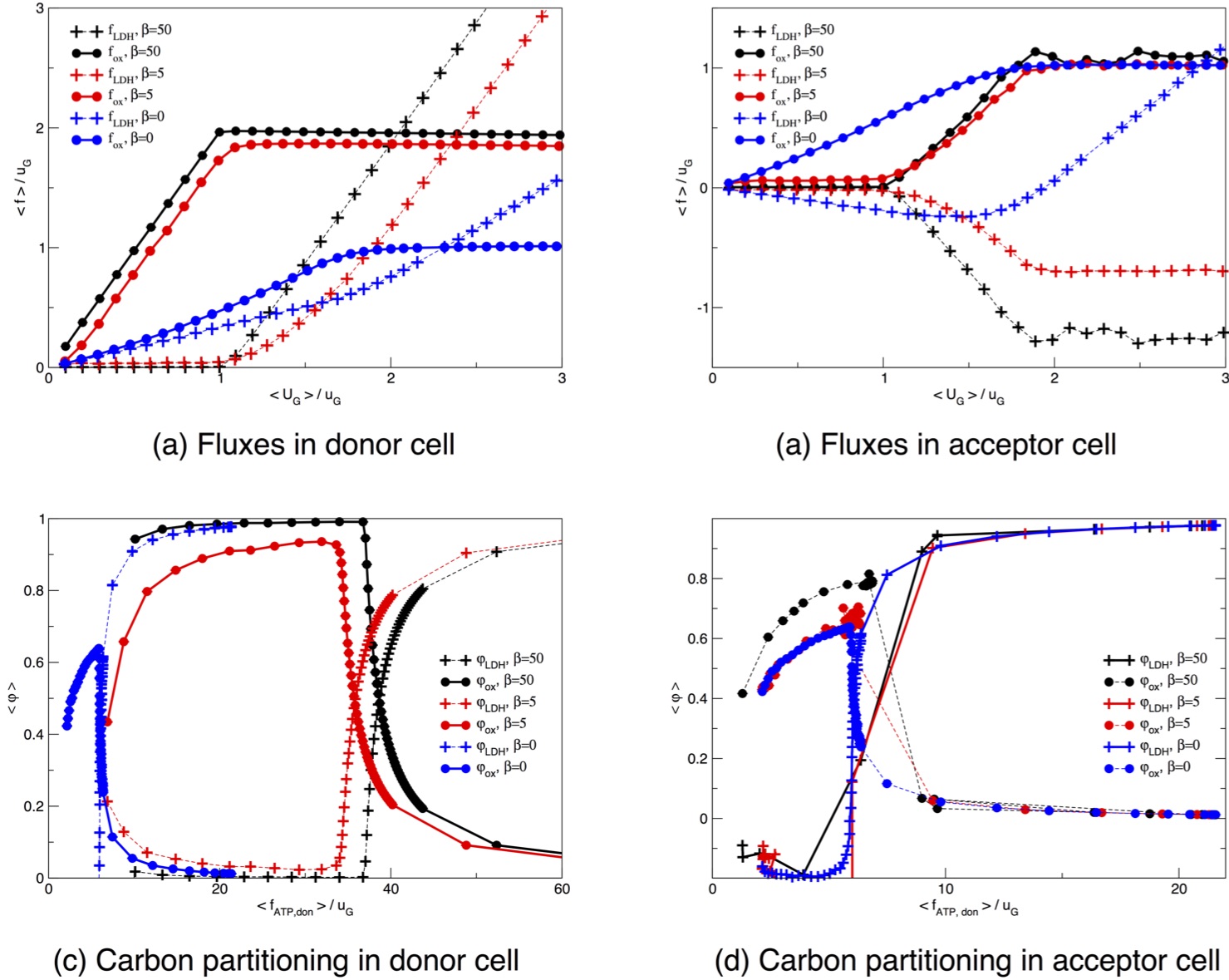}
\caption{\textbf{Oxidative and fermentative fluxes in a donor-acceptor system.} (a)--(b) Average flux through LDH (circles) and PDHm (crosses) as a function of the average glucose supplied to the two-cell system. 
(c)--(d) Fraction of ATP produced via glycolysis (circles) or via oxidative phosphorylation (crosses) as a function of the total ATP produced by the donor cell.
Pathways are normalized by the ``pyruvate-equivalent", i.e. by the sum of lactate and half-glucose intakes.
Curves describe the behaviour obtained for two coupled HCCN cells with an ATP-maximizing donor (black lines, $\beta=50$), a loosely maximizing donor (red lines, $\beta=5$) or for an unbiased sampling of the two-cell solution space (blue lines, $\beta=0$).
Error bars, which represents s.e.m., are smaller than symbol sizes.
\label{FIG: glucose distribution two cells}
}
\end{figure*}

We now consider two replicas of the HCCN and again distinguish between a lactate acceptor and a donor cell. We also impose that there is no external lactate source, and explore the scenario where the donor optimizes ATP production to a degree tuned by the parameter $\beta$. Fig.~\ref{FIG: 2HCCN GLUC ATP}a shows that, as expected, an ATP-maximizing donor cell sequesters all of the available glucose except for the small amount required for the acceptor's survival. 
This is reflected by the ATP production curves as a function of the total glucose supply displayed in Fig.~\ref{FIG: 2HCCN GLUC ATP}b. In such conditions, the acceptor's ATP production fluxes matches the minimum required for survival, i.e. $\atpmin$  (which is set to be equal to 1 $u_{\G}$ for simplicity) until the donor switches to aerobic glycolysis, thereby excreting lactate. The donor, on the other hand, goes through an initial phase of exclusive OXPHOS use, followed by a switch to aerobic glycolysis when the \rev{crowding} constraint does not allow for a further increment of the mitochondrial flux. In Fig.~\ref{FIG: glucose distribution two cells}a one indeed sees that glucose is mainly channeled to OXPHOS as long as the \rev{crowding} constraint allows for it. 
As soon as the latter is saturated, the donor diverts pyruvate to LDH, usefully regenerating NAD from NADH and thus avoiding glycolysis halt. LDH generates lactate, which is then expelled and intaken by the acceptor. As shown in Fig.~\ref{FIG: glucose distribution two cells}b, this is accompanied by a reversal of LDH in the acceptor: lactate is transformed into pyruvate that is then channeled to OXPHOS through PDHm. The acceptor can thus spare some cellular resources to produce pyruvate, at the cost of becoming strongly dependent on the donor for ATP production.

If one instead averages over all solutions without biasing for ATP production by setting $\beta=0$, the two cells  produce comparable amounts of ATP and share glucose more evenly (see Fig.~\ref{FIG: 2HCCN GLUC ATP}a). The fact that the donor turns out to employ slightly more glucose than the acceptor is due to the intrinsic asymmetry that is introduced by not letting the donor intake lactate.
This asymmetry is also sufficient to drive,
even when an unbiased statistical picture of the solution space is considered, a net lactate exchange as the typical behaviour
Both cells, however, produce a factor 6 (roughly) less ATP than the maximum possible, given the glucose influx. Despite a similar overall ATP production profile, the internal fluxes of donor and acceptor differ significantly. While the donor distributes   glucose almost evenly  through LDH and PDHm, the acceptor can only generate ATP via OXPHOS, since it intakes lactate as an extra source of carbon.
In Supplementary Fig.~S1,
we show that two symmetric cells with unbiased ATP production show identical glucose intake, ATP production, and internal flux distribution.
Even for symmetric cells, however, when the donor optimizes the ATP production, we observe lactate shuttling between donor and acceptor cells,
as illustrated again by 
Supplementary Fig.~S1.
In summary, for maximal ATP production ($\beta\to\infty$) there is no difference between ATP production and internal flux distributions of symmetric and asymmetric cells-couples. For simplicity in the analysis of the data, we used the asymmetric case where the donor cell cannot intake lactate.

We observe that the fraction of ATP produced via oxidative phosphorylation utilized by the acceptor cell differs substantially from the one of the donor cell. In Fig.~\ref{FIG: HCCN single cell}d we display the relative usage of the fermentative and the oxidative pathway and observe that, for low glucose supply, the acceptor cell only employs the oxidative pathway, independent of whether the donor cell maximizes ATP or not. When the donor cell maximizes ATP production, it sequesters most glucose and the acceptor cell is therefore forced to feed on lactate, which can only be usefully converted to ATP by means of OXPHOS. Conversely, when the donor cell does not maximize ATP production, it secretes a sizable amount of lactate because its metabolism is inefficient. The lactate acceptor, however, is also mainly feeding on glucose (see Fig.~\ref{FIG: 2HCCN GLUC ATP}a) and could in principle secrete lactate just like the donor cell. The reason why a purely oxidative metabolism is observed lies in the reversibility of LDH. Although lactate intake by the acceptor is small, it suffices to force the pyruvate flux towards TCA and OXPHOS, thereby making the acceptor more efficient than the donor in energetic terms.

\begin{figure}
\centering
\includegraphics[width=0.45\textwidth]{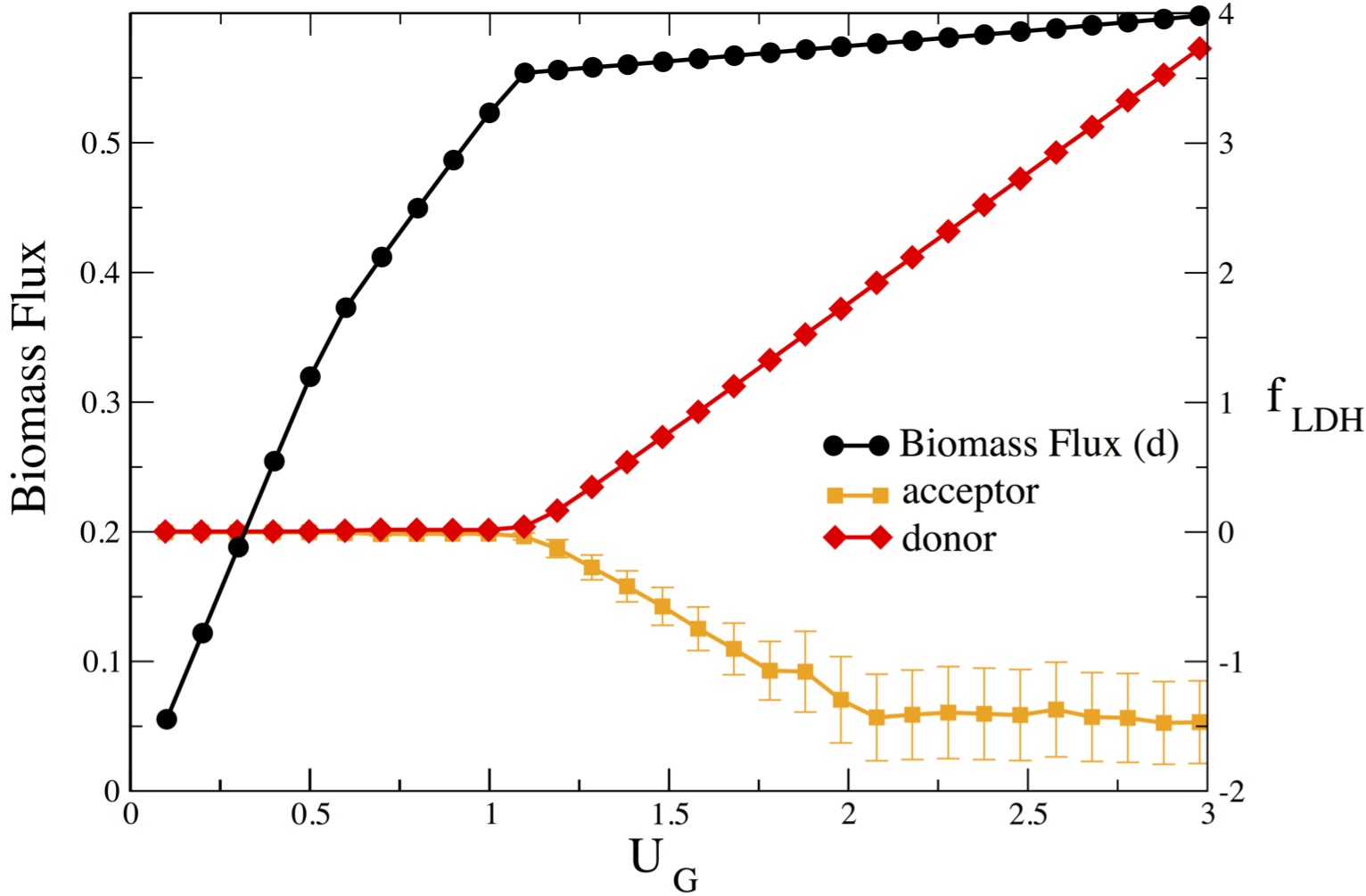}
\caption{\textbf{Donor's biomass flux and lactate overflow, and acceptor's lactate intake for two coupled HCCNs with a biomass-maximizing donor.} The qualitative behavior obtained when the donor is maximizing the ATP flux is reproduced in a more realistic case in which a biomass objective function is considered. See Table S9 for the biomass coefficients.}
\label{biomass}
\end{figure}

In order to assess the robustness of the occurrence of lactate overflow metabolism under the imposed constraints, we have further analyzed the flux configurations in a dono/acceptor system in which the donor maximizes the production of biomass precursors. We show here (see Fig. \ref{biomass}) results obtained by maximizing the biomass defined as in Table S9, which essentially reproduce those described above. 
Such a robustness becomes less surprising in the light of the fact that the qualitative features of the switch from oxidative to non-oxidative metabolism are obtained even in an unbiased sampling of the steady states.  This observation suggets that such a scenario is to a large degree embedded in the stoichiometry and in the main topological features of the underlying reaction network.

\rev{\subsection*{Correlation coefficients among metabolic fluxes of two cells coupled by lactate shuttle under glucose limitation}}

\begin{figure}
\centering
\includegraphics[width=0.33\textwidth]{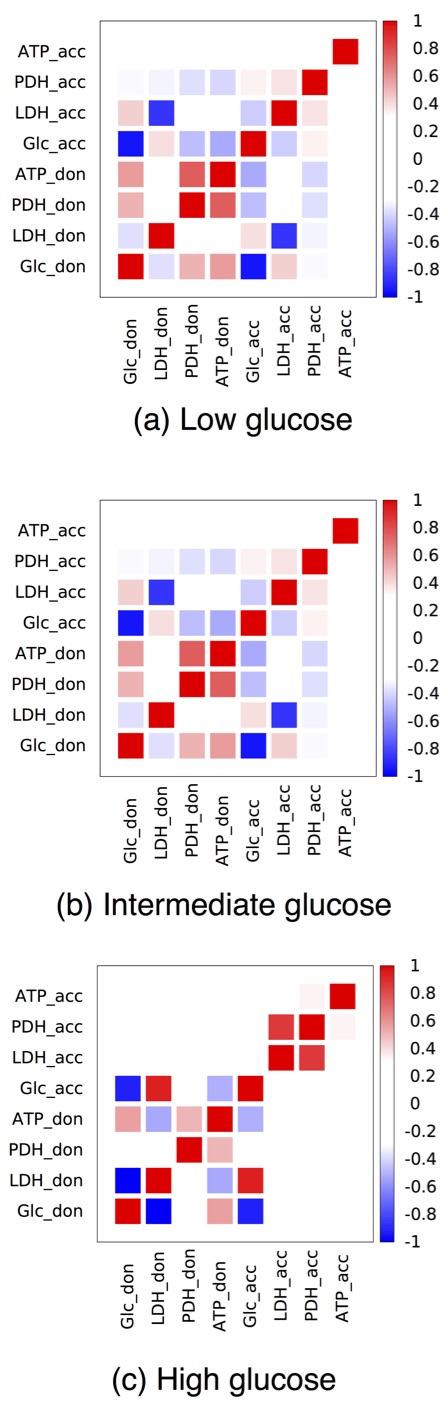}
\caption{\textbf{Restricted matrices of Pearson correlation coefficients for two coupled HCCN cells
when the lactate donor maximizes the ATP production and the overall
glucose supply is large shows that the two cells are not correlated.}
The intensity of the color represents the magnitude of the correlation coefficient (see scale on the right hand side). The two cells can independently access glucose and internal fluxes of the lactate acceptor and donor are essentially uncorrelated. The four representative reactions displayed for each cell are the glucose influx (Glc, a proxy for glycolytic activity), LDH (a proxy for lactate overflow and exchange), PDH (a proxy for oxidative metabolism) and the ATP production flux (ATP).}
\label{matrix}
\end{figure}

To formally assess the extent of metabolic coupling between an ATP-maximizing lactate donor and a lactate acceptor, we computed the matrix of normalized Pearson correlation coefficients of each pair of fluxes in the solutions sampled for different levels $U_\G$ of the glucose supplied to the system. The Pearson coefficient between random variables $X$ and $Y$ is defined as $r=\mathrm{cov}(X,Y)/(\sigma_X \sigma_Y)$, where $\mathrm{cov}(X,Y)$ denotes their covariance and $\sigma_X$ and $\sigma_Y$ stand for their respective standard deviations. $r$ ranges from $-1$ to $1$ and quantifies the linear dependence of the two variables. More precisely, the linear correlation between variables $X$ and $Y$ is more positive the closer $r$ is to $1$, and more negative the closer $r$ is to $-1$, while $X$ and $Y$ can be considered uncorrelated if $r\simeq 0$. For sakes of clarity, we have considered the correlations arising in a system formed by an ATP-maximizing lactate donor (large $\beta$) and a lactate acceptor. For smaller values of $\beta$ correlations get weaker while maintaing the same qualitative structure. Fig. \ref{matrix} displays three reduced correlation matrices (obtained for three different values of the overall glucose supply) where only a small subset of fluxes (each presentative of a different biologically relevant pathway) appears. Full matrices for three choices of $U_\G$ are instead shown in Figs. S3--S5.

Intuitively, correlation patterns should depend strongly on $U_\G$. The essence of this dependence, which clearly emerges from Fig. \ref{matrix}, is that cross-correlations between donor and acceptor build-up as the glucose supply increases, are maximal when the acceptor's energetics depends on donor-derived lactate, and then decrease again rapidly to zero when the glucose supply is large enough to sustain both cells.

If $U_\G$ is low (sub-threshold for lactate overflow), the ATP-maximizing donor \rev{necessarily} outcompetes the acceptor for the available glucose, establishing a degree of cross-correlations that is driven mainly by the unbalanced nutrient partitioning. The donor indeed employs OXPHOS almost exclusively, while still leaking out a small amount of lactate which is taken up by the acceptor (i.e. there is no accumulation in the tissue). This suffices to establish the overall correlation among cells that can be seen in Fig.~\ref{matrix}a.

As the glucose supply is increased further (above threshold for lactate overflow), the ATP-maximizing donor switches to aerobic glycolysis and secretes lactate that is shuttled to the acceptor in large amounts (viz. the strong negative correlation arising between the donor's lactate outflux and the acceptor's lactate influx). In such conditions, the overall cross-correlations increase, see Fig.~\ref{matrix}b. In particular, glycolytic fluxes in donor and acceptor anti-correlate, oxidative pathways weakly anti-correlate, while te donor's biosynthetic pathways correlate weakly with the acceptor's oxidative pathways. Notice also that ATP flux in the donor weakly anticorrelates with every pathway in the acceptor's metabolism.

At high enough $U_\G$, finally, cross-correlations decay as both cells can rely entirely on glucose for their energetics. Because the maximum glucose uptake for an HCCN is given by $U_{\G}^{\max}=\Phi_{\ATP}/\left(a_{\G}+2a_{\L}\right)$ (this is determined by the \rev{crowding} constraint  \eqref{eq: HCCN volume constraint} when glucose is entirely channeled to fermentation, i.e. when $f_{\HEX}=U_{\G,\D}=f_{\LDH}/2$), one may assume that  $U_{\G}$ is ``large enough" when it is larger than $2 U_\G^{\max}$, as both cells would in this case have access to as much glucose as they can intake. In this case, no donor-acceptor correlations should be expected. Fig.~\ref{matrix}c shows that this is indeed the case: intra-cellular correlations, described by the diagonal blocks, greatly exceed inter-cellular ones (off-diagonal blocks). 

In summary, when glucose availability is limited, the fast-growing donor \rev{necessarily} outcompetes the acceptor for glucose and a network of cross-correlations is established between the metabolic systems of the two cells. Metabolic pathways in donor and acceptor correlated however most strongly when a lactate exchange from donor to acceptor is established, and decay again as the two cells become effectively decoupled in glucose-rich media. The build-up of a metabolic partnership is perhaps best seen by the fact that the donor's glucose influx $U_{\G,\D}$ develops a correlation with the acceptor's LDH flux at intermediate values of $U_\G$. Note also that $U_{\G,\D}$ correlates more and more strongly with the donor's LDH flux as $U_\G$ increases.

\bigskip

\section*{Discussion}

\subsection*{Experimental evidence on lactate shuttles in cancer}

While lactate shuttling is a likely scenario in many physiological conditions, from muscle cells undergoing intense activity \cite{Brooks:2009p4313} to neuron-glia energetics \cite{Barros:2013p4310,Belanger:2011p4279, Dienel:2012p4280, Massucci:2013p4278}, and has been reported to take place intra-cellularly at the mitochondrial and peroxisomal membrane \cite{Brooks:2009p4313},  no direct measurement of inter-cellular lactate exchange in cancer exists. Strong indirect evidence, however, suggests that such a scenario is indeed plausible.

Indeed, tumor-stroma and tumor-tumor \rev{metabolic} interactions are currently being characterized at various levels. The emerging picture increasingly suggests that cancer sustainment is a non-cell-autonomous phenomenon \cite{pmid11737888,pmid25079331,Zhang:2012p4006,Choi:2013p4423} and that stromal cells might be potential targets for cancer treatment \cite{pmid25259922}. Lactate exchange in particular has been investigated both as fueling oxidative cancer cells and as supporting non-aberrant cancer-associated cells \cite{Draoui:2011p3948,Doherty:2013p4183}. The latter case corresponds to the model discussed here. Signatures of tumor-to-stroma lactate shuttle have been reported in terms of the over-expression of monocarboxylate lactate transporters jointly with increased PDH activity in cancer-associated fibroblasts (CAFs) \cite{pmid15736311,pmid16423989}. Likewise, tumor-derived lactate has been found to play a major signaling role (specifically, for the initiation of tumor angiogenesis) in vascular endothelial cells \cite{pmid21300765}. Inhibition of the lactate transporter MCT1 has therefore been proposed as a possible anti-angiogenic strategy \cite{Sonveaux:2008p3954}. 

Taken together, these observations suggest that tumor cells and their associated stromal and vessel cells can be seen as a collective, synergistic metabolic unit where each compartment carries out complementary functions reflected in their energetic strategies. In such a microenvironment, aerobic stromal cells serve an essential `bioremediative' role by removing potentially toxic compounds, thereby reducing acidity and positively feeding back on anaerobic cancer growth. The existence of a consistent imbalance in energetic demands across different compartments is crucial to establish this scenario.

Intercellular shuttling of lactate towards oxidative cancer cells is known to occur in two distinct forms, namely either from non-oxidative (e.g. hypoxic) to oxidative tumor cells \cite{Sonveaux:2008p3954} or through the  `reverse Warburg effect', i.e. the onset of aerobic glycolysis in CAFs \cite{pmid21300172,pmid24597899,pmid22850421,pmid24886074,WhitakerMenezes:2011p4016}. The former case describes the metabolic sysmbiosis that is established e.g. between more and less hypoxic regions of a tumor, which putatively allows glucose to be delivered more efficiently to more hypoxic regions \cite{pmid21704401} (see \cite{ardeshir} for a game-theoretic approach to modeling this kind of tumor-tumor interplay). In the latter scenario, cancer cells induce oxidative stress in CAFs, resulting in CAFs switching to anaerobic glycolysis \cite{pmid21778829}. The secreted lactate is then imported by cancer cells for use in aerobic pathways. This effect has indeed been observed {\it in vitro} \cite{pmid22850421,WhitakerMenezes:2011p4016}, and its relevance {\it in vivo} is currently under scrutiny \cite{pmid24597899}. Clearly, however, this type of shuttling is not necessarily triggered by a strong imbalance in energetic demands between the lactate donor and acceptor cells, and different constraints in metabolic activity are likely to be required to understand its origin within genome-scale models.

It is worth noting that recently developed microfluidic platforms allow to probe the tumor microenvironment with unprecedented resolution and control over nutrient supply \cite{methods}. More light will hopefully be shed on its activity as a functional metabolic assembly and on the role of lactate in specific.

Going beyond cancer, however, such studies highlight the fact that replacing the natural environment in which cells live with a cell culture that is possibly optimized for the cell's needs might severely limit our ability to understand the actual behavior of the system {\it in vivo}. On the other hand, they suggest that cell-autonomous models may be unable to capture some essential features of the energetics of cells: when cells employing different energetic demands share a limited resource, a mutually beneficial microenvironment characterized by a large-scale exchange of chemical species can be established. A thorough understanding of cellular growth strategies should take these aspects into account.

\subsection*{Conclusions}

In this article we have studied the lactate-driven coupling that is established between cells with different energetic requirements, showing that a lactate shuttle appears robustly as a consequence of basic physico-chemical  constraints. \rev{Our model covers time scales over which the cellular metabolic networks have to adapt to the establishment of an imbalance in energy demands (as well as, for instance, in the levels of main carbon transporters), which are much shorter than the time required to alter the nutrient availability profile significantly (e.g. via vascularization).} We have illustrated the emergent \rev{metabolic} interaction scenario in a simplified model that captures the essential fact that, for two such cells, recycling fermentation products may be a good strategy to optimize cellular glucose usage. To reach these conclusions, we have considered a human metabolic network model of cells interacting via lactate shuttle and sampled its feasible metabolic states according to a prescribed probability distribution by which we can tune the degree to which one of the two cells is pushing its ATP production. The scenario obtained in a highly simplified (but exactly solvable) model is fully retrieved in this case, where a thorough analysis of correlation patterns reveals further details of the cell-to-cell coupling.

Such a coupling would have serious physiological implications. First, lactate recycling implies that lactate may accumulate in tissues only in late-stage tumors. Before neighboring cells saturate their processing capabilities, the lactate expelled by cancer cells would be taken in by non-aberrant cells, suggesting that localized lactate measurements might identify early-stage tumors. Second, lactate shuttling and recycling systems should be regarded as potential targets for treatment. 

There are many aspects of the approach presented here that can be further dissected. First is the nature of constraints. From a physical point of view, imposing ``volume'' constraints of the type considered here is equivalent to imposing finite flux capacities on energy-efficient pathways. There are however strong indications that unicellular organisms actively down-regulate energy-efficient pathways at high nutrient levels \cite{Lemuth:2008p4288, DeRisi:1997p4290}. The fact that cells invest energy and resources in silencing the synthesis of enzymes involved in aerobic pathways suggests that the switch to an energy-inefficient pathway provides a real, physiological advantage for the cell. Its origin and nature are currently not understood, although the inclusion of additional ``costs'' due to regulation in metabolic models strongly suggests that the observed phenomenology can be captured, at least to some degree, by accounting for the fact that a shift in energetic strategy requires a sizeable change of a cell's protein repertoire \cite{Molenaar:2009p3978}.

It is also important to note that the ETC in eukaryotes takes place in a separate cellular compartment (the mitochondrion), at odds with prokaryotes like bacteria, where it occurs is the cell's periplasm. We have entirely neglected the complications due to spatial organization, which is reasonable as long as one wants to focus on the ATP yield of energy-producing strategies. However, this aspect will become important for higher-resolution modeling at genome scale. \rev{Likewise, a more refined spatial modeling will allow for an in silico analysis of a possible role of intra-cellular lactate shuttling in cancer \cite{Brooks:2009p4313}. In brief, the key idea behind this is that glycolysis-derived lactate could be employed directly as an additional energy source for mitochondrial oxidation instead of being excreted. Elementary biochemical properties of the LDH reaction indeed would appear to favour lactate (rather than pyruvate) production in the cytosol. Moreover, experimental facts indicate a possible role of intra-cellular lactate shuttling in cardiomyocites, neurons and skeletal muscle cells facing high energy demands \cite{Cruz:2012p4421}. Accounting for it in a computational scheme may help reveal novel details about possible pathways of energy production during physical exercise or in rapidly growing cells.}

\rev{Furtermore, while this work has focused on lactate as a key mediator of metabolic, energy-driven intercellular interactions, it is worth stressing that recent studies focusing on the comprehensive analysis of the exometabolome of growing cells has revealed that metabolic interactions encompass many more chemical species and is tightly coupled with the growth regime. In bacteria, for instance, released intermediates include not only carbon equivalents (like acetate, pyruvate and ethanol) but also amino acids and central metabolic intermediates [like fructose-6-phosphate, glucose-6-phosphate, 2/3-phosphoglycerates, phosphoenolpyruvate, acetyl-CoA, citrate and $\alpha$-ketoglutarate], higher amounts of which have been found to be consistently released when carbon availability is high, while intermediates are typically in-taken in carbon-limitation \cite{Paczia:2012p4266}. Similar interactions are now known to occur in cancer: leukaemia cells have indeed been recently shown to employ cysteine derived from bone-marrow stromal cells as a means to fight oxidative stress \cite{Zhang:2012p4006}. As the scenario underpinning the establishment of such couplings is increasingly elucidated, it will become possible to set up more refined and realistic models to capture a greater extent of their physiological relevance.}

\rev{Finally, it should be kept in mind that aerobic glycolysis in cancer cells, especially at later stages of oncogenic development, could be due to the fact that the tumor microenvironment is hypoxic. The model we consider here addresses a faster time scale, over which the energy balance of neighbouring cells disrupts while the overall amount of resources available remains (roughly) unchanged, and suggests that at these stages tumor-to-stroma metabolic interactions can provide a mutually beneficial solution to the imbalance. It is also worth mentioning that cases are known in which malignant cells rely on increased oxidative phosphorylation rather than on aerobic glycolysis for energy production (as found, for instance, in transformed human mesenchymal stem cells \cite{Funes:2007p4268}). Taken together, these results suggest that cancer's bioenergetics, starting with the aerobic glycolysis versus oxidative phosphorylation ``dilemma'', may be a dynamically modulated process that differs widely across tumor types \cite{Jose:2011p4422}}

The modeling approach discussed here, based on exploring the solution space induced by mass-balance equations rather than on optimizing a prescribed objective function, is in our view the most suited to deal with multi-cell systems in which extended cell-to-cell \rev{metabolic} interactions are established by one cell's deregulated metabolic demands. The sampling method employed here is scalable and easy to implement, providing a highly promising tool for further studies. It would in particular be important to analyze the emergence of these effects  in full-fledged genome-scale models of specific cancers, as can be obtained e.g. by refining reconstructions of human metabolic networks with cancer expression data.
The full complexity of metabolic cell-to-cell interactions has only just started to be uncovered. Understanding its functional relevance by detailed {\it in silico} models may provide new insight into the mechanism of cancer progression and further advance the search of specific, targeted treatments.
\rev{It would be important to assess the relevance of lactate shuttling in the context of tumors through experiments probing metabolism (or metabolite exchanges) directly, rather than by collecting indirect evidence in the up- or down-regulation of specific transporters, and metabolic solution space sampling is known to provide useful keys for experiment design \cite{Price:2004p4298}. Our results indicate at least two potential ways to obtain useful information. One possibility requires setting up an assay where the ability of non-aberrant cells to intake lactate can be modulated in the presence of bona-fide lactate-secreting cancer cells. Such a modulation could be achieved, for instance, by either varying the relative concentration of donor- and acceptor-cells, or by changing the expression of lactate and glucose transporters in supporting cells. It is reasonable to expect that, in such a setup, the extracellular acidification rate due to the accumulation of lactate in the extracellular medium should negatively correlate with the lactate removal capacity of the supporting cells. On the other hand, results from our statistical sampling of the solution space suggest that different pathways (and hence, likely, the expression levels of different enzymes) should be tuned in a specific way by changing the overall glucose supply in a mixed culture shared by aberrant and non-aberrant cells. Monitoring expression levels in different, controlled nutrient conditions would therefore provide key validation to the coupling picture discussed here (and, more generally, to any cell-to-cell coupling scenario for cancer sustainment).}

\section*{Materials and methods}

\subsection*{Human core catabolic network \rev{for a single cell}}

We built a realistic metabolic network of ATP production from the decompartmentalized human reactome Recon-1 \cite{Duarte:2007p3938} \rev{that we use throughout the manuscript}. (While the more recent Recon-2 reactome \cite{Thiele:2013p4010} provides a higher resolution reconstruction of human metabolism, the degree of detail provided by Recon-1 more than suffices for our purposes.) The HCCN includes the following Recon-1 pathways: Glycolysis, Pentose Phosphate Pathway, Citric Acid Cycle, and Oxidative Phosphorylation.  In addition, we included anabolic pathways for the production of biomass precursors (amino acids and fatty acids) in an effective way. Details of the reconstructed network are given in Supporting Text S1. We then proceeded by removing the leaves of the resulting network (a necessary pre-processing step required to implement the mass-balance constraints described below). To rid the model of {\it a priori} infeasible loops, we resorted to the method described in \cite{DeMartino:2013p4115}. A single infeasible cycle was identified, which was fixed by leaving only one of the two isoforms of the isocitrate dehydrogenase that uses NADP as cofactor.

The final HCCN we employed in this study is composed of 67 chemical species (listed in Table~S2) and 75 reactions (listed in Tables~S3 and S4), including the uptake/secretion of 11 metabolites (namely molecular oxygen, carbon dioxide, water, hydrogen, lactate, glucose, ammonia, phenylalanine, methionine, glutamine, and methyl group---which represents a generic methylation--whose bounds are fixed as described in Table~S5), and reactions consuming ATP, aminoacids and palmitic acid. Among intracellular reactions, $23$ are reversible according to Recon-1's thermodynamic assignments. We have considered a medium with variable maximum glucose intake, fixed maximum glutamine intake and unbounded oxygen availability. Finally, ATP consumption presents a lower bound standing for the minimum ATP consumption flux $\atpmin$ necessary for cell survival (the bounds of all reactions are listed in Table~S6 and S7).

\rev{\subsection*{Human core catabolic network for two coupled cells}}

In order to analyze the coupling of two identical cells, we simply replicated the HCCN twice, the only difference lying in the uptake reaction for lactate. In specific, the lactate donor can only secrete lactate (through an irreversible reaction), while the acceptor can both excrete and import it. The coupling is ultimately determined by the partitioning of glucose and by the shuttling of lactate. The former is a shared resource, while the latter can be exchanged between the two cells: both cells can produce lactate, but there is no external source of lactate. We have included two further constraints for  the sum of glucose and lactate flux of the donor-acceptor system, i.e., $U_{\G,\D}+U_{\G,\A} \leq U_{\G}$ and $U_{\LAC,\D}+U_{\LAC,\A}\ge0 $ (see Table~S8). As for a single HCCN, glucose can only be imported while lactate can only be excreted. 

\rev{\subsection*{Crowding constraint for the human core catabolic network}}

The \rev{crowding} constraint within the HCCN has been implemented as 
\begin{equation}
a_{\GL}f_{\HEX}+a_{\M}(f_{\PDH}{+ f_{\GLUN}})+a_{\L}f_{\LDH}\leq\Phi_{\ATP}\label{eq: HCCN volume constraint SM}
\end{equation}
where $f_{\HEX}$ denotes the Hexokinase-1 flux, $f_{\PDH}$ denotes the flux through pyruvate dehydrogenase, by which pyruvate is diverted into the TCA cycle, and $f_{\GLUN}$ denotes the flux through glutaminase. For $a_{\GL}$, $a_{\L}$ and $a_{\PDH}$ we use the numerical values given in \cite{Vazquez:2010p3868} for a coarse-grained model, since for unitary flux both $f_{\HEX}$ and $f_{\M}$ transport the same amount of carbons present in glucose, which is exactly what the effective fluxes defined in \cite{Vazquez:2010p3868} do.

Inequality \eqref{eq: HCCN volume constraint SM}  is valid for a cell that excretes lactate. When lactate can also be intaken, as occurs for the coupled HCCN cells, the flux through LDH can become negative. The more general form of the \rev{crowding} constraint we consider is therefore
\begin{equation}
a_{\GL}f_{\HEX}+a_{\M}(f_{\PDH}+{f_{\GLUN}})+a_{\L}|f_{\LDH}|\leq\Phi_{\ATP}.\label{eq: HCCN volume constraint abs val}
\end{equation}

For both the single HCCN and the two-HCCN models, we provide \rev{as supplementary materials} files containing the lists of metabolites and reactions (with bounds), and the explicit expressions of each reaction in the HCCN in terms of independent variables, 
which provide a full characterization of the solution space polytope. The files can also be downloaded from http://chimera.roma1.infn.it/SYSBIO, where an SBML file with the model and a text file with the stoichiometric matrix of the single cell HCCN are also made available. \rev{HCCN for a single cell has also been deposited in BioModels Database (biomodels.org) and assigned the identifier MODEL1506170000.}


\bigskip

\subsection*{\rev{Representation of the solution space of constraint-based metabolic networks as a convex polytope}}

In mass-balance setups, a set of reaction fluxes $\mathbf{f}=\{f_r\}$ ($r=1,\ldots, N$) describes a non-equilibrium stationary state if it satisfies the conditions
\begin{equation}\label{mbe}
\sum_{r=1}^N S_{rm} f_{r}=0 \quad,\quad m=1,\dots, M~~,
\end{equation}
with prescribed bounds $f_r^-\leq f_r\leq f_r^+$. (Clearly, the zero flux vector is excluded from these considerations.) Here, $S_{rm}$ denotes the stoichiometric index of chemical species $m$ ($m=1,\ldots M$) in reaction $r$. From a geometric viewpoint, (\ref{mbe}) defines a convex polytope, while every point inside it represents a feasible flux configuration. In absence of criteria allowing to select specific flux vectors from the solution space (e.g. via the maximization of an objective function), uniform sampling provides important information about average fluxes and correlations that not only describe the viable flux patterns in statistical terms, but may also give relevant guidelines for designing experiments \cite{Schellenberger:2012p4302}. When the dimension of the solution space is sufficiently small (typically around 10), Monte Carlo sampling, including straightforward rejection methods \cite{Price:2004p4298}, can be applied. At higher dimensions, instead, exceeding computing times require the use of more advanced techniques.

\subsection*{\rev{Dimensional reduction of the solution space}}

We are interested in sampling the solution space of (\ref{mbe}) with $\mathbf{S}=\{S_{rm}\}$ given by the HCCN with the prescribed probability measure defined in (\ref{gibbs}).
Mass balance conditions constrain many of the  75 fluxes to be dependent on other variables. To explore efficiently the space of viable flux configurations, it is convenient to solve all such dependencies analytically and then sample the (much smaller) space spanned by independent fluxes.  To begin with, we transformed $\mathbf{S}$ to its Reduced Row Echelon Form (RREF) through
Gauss-Jordan elimination \rev{, which can be implemented by standard packages.}
Because the RREF of any matrix is unique, the RREF of the stoichiometric matrix applied to the reaction network uniquely represents the dependent fluxes as a function of the independent ones. The single-HCCN model turns out to have only  17 independent variables, while the two-HCCN version has 34.
These numbers also represent the dimensions $D$ of the respective solution space polytopes.

\subsection*{\rev{Hit-and-run method allows to sample convex polytopes efficiently}}

The Hit-and-Run (HR) method \cite{Smith:1996p4127, Turcin:1971} can uniformly and efficiently sample a convex polytope, provided one starts from a point inside it. The problem of finding such a point can be easily solved, e.g., via Motzkin relaxation \cite{Motzkin:1954p4304, DeMartino:2013p4115}. In brief, HR builds a Markov chain in two steps. First, starting from a point inside the polytope, a random direction is extracted. Second, along this direction a new point is chosen uniformly at random inside the polytope. The scheme is then iterated starting from the new internal point. HR is efficient because once the random direction is generated the search space is reduced to a segment. In our implementation, we generate the random direction using the Marsaglia-Bray method \cite{Marsaglia:1964p4307}. To pick a point at random along the segment,  instead, we first compute the two intersections with the polytope and then extract a uniform random number inside the segment identified by these two points on the line.

\subsection*{\rev{The Lovasz method speeds up the sampling of heterogeneous polytopes}}

The above procedure, while fully exact, may however present a drawback in concrete cases. In fact, the mixing time of HR in convex domains scales as $\mathcal{O}[D^{2}(R/r)^{2}]$ Monte Carlo steps, where  $D$ is the spatial dimension of the polytope while $r$ and $R$ are, respectively, the radius of the maximum inscribed hypersphere and of the minimum circumscribed hypersphere \cite{Lovasz:1999p4121}. It is clear that the $R/r$ factor can increase convergence times dramatically for strongly heterogeneous polytopes. This is precisely the case for the  space of feasible steady states in a large-scale metabolic network, where flux distributions can be so heterogeneous as to span 5 orders of magnitudes \cite{Almaas:2004p4295, Braunstein:2008p4296, Martelli:2009p3766}. For the network analyzed here, the ratio between the ellipsoid's longest and smallest axes turned out to be of order  $10^{4}$, leading to a condition number of order $R^{2}/r^{2}\simeq10^{8}$, a value that would render straightforward HR too expensive in terms of CPU time. To circumvent this ill-conditioning problem, we have resorted to a standard pre-processing step that identifies the ellipsoid that best approximates the shape of the polytope \cite{lovazsbook}. Extracting vectors uniformly on such an ellipsoid guarantees that the directions along the longer axis are chosen more often, thereby removing ill conditioning. In quantitative terms, such a pre-processing gives HR a mixing time of $\mathcal{O}[D^{7/2}]$ Monte Carlo steps, thus implying a fully polynomial scaling with $D$. After pre-processing the slowest HR mode was found to decorrelate in an affordable time of $\mathcal{O}(10^{4})$ Monte Carlo steps. Furthermore, the overall pre-processing time was negligible compared to the time spent for the actual solution space sampling, while the center of the ellipsoid provides an excellent starting point for the algorithm. For more details of this method we refer the reader to \cite{DeMartino:2015p4294}. 

\subsection*{\rev{A method for tuning the optimization of a given metabolic flux for constrained based metabolic networks}}

HR is easily modified to sample points according to any given flux distribution $P(\mathbf{{f})}$ by  imposing that the selection of points along the segment takes place according to how the distribution $P$ projects onto the chosen direction rather than uniformly.
To modulate $P$ via a linear function $L(\mathbf{{f})}$ of the fluxes, it is convenient to use the Boltzmann distribution $P(\mathbf{{f})}\propto\exp{[\beta L(\mathbf{{f}})]}$, where $\beta\geq 0$ is an interpolation parameter. By varying $\beta$, one passes from an uniform sampling (corresponding to $\beta=0$) to sampling flux configurations that maximize $L(\mathbf{{f})}$ (corresponding to $\beta\gg 1$). For the HCCN network studied here, the ATP production $f_{\ATP}$ is maximized using the above method with $P(\mathbf{f})\propto \exp\left[\beta f_{ATP}\right]$. The value of $\beta$ for which $f_\ATP$ is effectively maximized can be determined numerically. In Fig.~S6 we plot the ATP production as a function of the $\beta$ for five different values of the maximal glucose available for two coupled HCCN cells. One sees that optimal $f_\ATP$ is already achieved when $\beta\simeq 50$.

\rev{
To produce the graphs presented in Figs.~\ref{FIG: HCCN single cell} through \ref{matrix} and S1 through S6, we applied the Lovasz method to the reduced row echelon forms of the HCCN networks and sampled the corresponding regularized polytopes by means of the Hit-and-run method. For curves with $\beta\ne0$, we used the Boltzman weight presented above. The apply the Lovasz method and the uniform and optimized HR sampling, we used C++ programs that we developed and that can be freely
downloaded at http://chimera.roma1.infn.it/SYSBIO.}

\section*{Author contributions}
F.C. and D.D.M. performed the experiments. All authors conceived and designed experiments, analyzed the data, and contributed to the writing of the manuscript.

\section*{Acknowledgments}
Work supported by the DREAM Seed Project of the Italian Institute of Technology, by the joint IIT/Sapienza Lab “Nanomedicine”, and by the PRIN project “Statistical mechanics of disordered complex systems” of the Italian Ministry of University and Research. The IIT Platform “Computation” is gratefully acknowledged.

\newpage

\widetext

\section*{Supplementary Text}

\subsection*{Text S1. Human Catabolic Core Network: biosynthetic reactions}

Besides the major central carbon metabolic pathways, i.e. Glycolysis, Pentose Phosphate Pathway, Citric Acid Cycle, and Oxidative Phosphorylation, our model includes
most reactions of Glutamate metabolism as well as lumped reactions describing the biosynthesis of all non-essential amino acids (i.e. the amino acids that can be synthesized by human cells) and palmitate, a key precursor of other fatty acids. 

Amino acid biosynthetic processes can be divided roughly in two classes. The synthesis of tyrosine and cysteine requires respectively the essential amino acids phenylalanine and methionine. All other amino acids can instead be synthetized from glutamine plus metabolites from the central carbon pathways. The processes leading to amino acid synthesis are of various types, from simple single transaminase reactions to a complex chain of enzymatic reactions. In the latter case, by means of mass balance arguments we can lump together individual reaction steps to obtain effective reactions whose net output corresponds to the desired amino acid(s).
In all cases, our starting points were the reactions and pathways included in the reference metabolic network reconstruction Recon-1~[21].
In the \texttt{.xml} file provided as Supporting Material, we include the list of the corresponding Recon-1 reactions that constitute the
full pathway.  In the following, we briefly outline these effective reactions.
\begin{itemize}
\item Glutamate synthesis: from glutamine via glutaminase (GLUN):
\begin{equation}
{\rm gln\_L} + \textrm{H}_2\textrm{O} \to {\rm glu\_L} + \textrm{NH}_4
\end{equation} 
\item Alanine and aspartate synthesis: from glutamate through the transaminase reactions
\begin{gather}
{\rm akg} + {\rm ala\_L}  \leftrightarrow {\rm glu\_L} + {\rm pyr}\\
{\rm akg} + {\rm asp\_L} \leftrightarrow {\rm glu\_L} + {\rm oaa}
\end{gather}
(akg = $\alpha{\rm \text{-}ketoglutarate}$ ; pyr = pyruvate ; oaa = oxaloacetate)
\item Proline synthesis: from glutamate via glutamate 5-kinase (GLU5K), glutamate-5-semialdehyde dehydrogenase (G5SD), L-glutamate 5-semialdehyde dehydratase (G5SAD) and pyrroline-5-carboxylate reductase (P5CR):
\begin{equation}
\textrm{glu\_L} + \textrm{ATP} + (2) \textrm{NADPH} + (2) {\rm H}^+ \to \textrm{pro\_L} + \textrm{ADP} + \textrm{Pi} + \textrm{H}_2\textrm{O} + (2) \textrm{NADP} 
\end{equation}
\item Serine synthesis: from glutamate via phosphoglycerate dehydrogenase (PGCD), phosphoserine transaminase (PSERT) and phosphoserine phosphatase (PSP):
\begin{equation}
{\rm 3pg} + \textrm{NAD} + \textrm{H}_2\textrm{O} + {\rm glu\_L} \to  {\rm ser\_L} + {\rm akg} + \textrm{NADH} + \textrm{Pi} + {\rm H}^+
\end{equation}
(3pg = $3{\rm \text{-}phosphoglyceric \,acid}$)
\item Glycine synthesis: from serinevia glycine hydroxymethyltransferase (GHMT), methylenetetrahydrofolate 
dehydrogenase (MTHFD), methenyltetrahydrofolate cyclohydrolase (MTHFC) and formyltetrahydrofolate dehydrogenase (FDH):
\begin{equation}
{\rm ser\_L} + \textrm{H}_2\textrm{O} + (2) \textrm{NADP} \to {\rm gly} + {\rm CO}_2 + (2) \textrm{NADPH} + (2) {\rm H}^+
\end{equation}
 \item Arginine synthesis: from glutamate and aspartate via glutamate 5-kinase (GLU5K),  glutamate-5-semialdehyde dehydrogenase (G5SD), ornithine transaminase (ORNTA), carbamoyl-phosphate synthase (CBPS), ornithine carbamoyltransferase (OCBT), argininosuccinate synthase (ARGSS) and argininosuccinate lyase (ARGSL):
\begin{multline}
{\rm asp\_L} + {\rm glu\_L} + {\rm gln\_L} + (5) \textrm{ATP} + \textrm{NADPH}  +(3) \textrm{H}_2\textrm{O} + \textrm{CO}_2 \to \\ 
\to {\rm arg\_L} + {\rm fum} + {\rm akg} + (5) \textrm{ADP} + (5) \textrm{Pi} +  \textrm{NADP}+ (5) {\rm H}^+
\end{multline} 
\item Asparagine synthesis: from aspartate and glutamine via asparagine synthase (ASNS): 
\begin{equation}
{\rm asp\_L} + (2) \textrm{ATP} + {\rm gln\_L} + (2) \textrm{H}_2\textrm{O} \to (2) \textrm{ADP} + {\rm asn\_L} + {\rm glu} + (2) {\rm H}^+ + (2) \textrm{Pi}
\end{equation} 
\item Cysteine synthesis: from serine and methionine via methionine adenosyltransferase (METAT), adenosylhomocysteinase (AHC), cystathionine beta-synthase (CBS), cystathionine gamma-lyase (CSE), 2-Oxobutanoate dehydrogenase, Propionyl-CoA carboxylase (PCCA) and methylmalonyl-CoA mutase (MMM):
\begin{multline}
{\rm met\_L} + {\rm ser\_L} + {\rm coa} + (4) \textrm{ATP} + \textrm{NAD}  + (4) \textrm{H}_2\textrm{O}  \to \\  \to{\rm cys\_L} + {\rm succoa} + (4) \textrm{ADP} + (4) \textrm{Pi} + (4) {\rm H}^+ + \textrm{NADH} + \textrm{NH}_4  + \textrm{CH}_3
\end{multline}
(coa = coenzyme A ; succoa = succinyl-coa)
\item Tyrosine synthesis: from phenylalanine via L-Phenylalanine hydroxylase, tetrahydrobiopterin oxidoreductase, tetrahydrobiopterin-4a-carbinolamine dehydratase and dihydropteridine reductase:
\begin{equation}
{\rm phe\_L} + \textrm{NADH} + {\rm O}_2 + {\rm H}^+  \to  {\rm tyr\_L} + \textrm{NAD} + \textrm{H}_2\textrm{O}
\end{equation} 
\item Palmitic acid synthesis:
\begin{multline}
(8) {\rm accoa} + (23) \textrm{ATP} + (14) \textrm{NADPH} + (17) \textrm{H}_2\textrm{O} \to \\ \to
{\rm palmitate} + (8) {\rm coa} + (14) \textrm{NADP}  + (23) \textrm{ADP}  +(23) \textrm{Pi}  + (10) {\rm H}^+ 
\end{multline}
(accoa = acetyl-coa)

\end{itemize}
In addition, we included the reactions for superoxyde reduction and ${\rm FADH}_2$ oxidation, that are necessary to correctly account for the redox state of the cell. For sakes of simplicity, we lumped together the reactions catalyzed by superoxide dismutase, 
gluthatione peroxidase 
and glutathione reductase 
to the single reaction 
\begin{equation}
{\rm NADPH}+{\rm O}_{2}^{-}+2{\rm H}^{+}\rightarrow {\rm NADP}+2{\rm H}_{2}{\rm O}~~.
\end{equation}
Likewise, we unified the two reactions that oxidate FADH$_2$, thereby reducing ubiquinone by means of the Electron transport flavoprotein-ubiquinone oxidoreductase (ETF) to
\begin{equation}
{\rm FADH}_{2}+{\rm Q10}\rightarrow {\rm FAD} + {\rm Q10H}_{2}~.
\end{equation}

\subsection*{Text S2. Maximizing the flux of precursors: results}

In the main text, we show that lactate overflow and shuttle arises under both biomass and ATP maximization.
Here, we show the metabolic flux patterns obtained when in the lactate donor the production of single biomass precursors are maximized, to assess the robustness of the
lactate overflow and shuttle.
To limit the amino acid productions, we have assumed a fixed maximum total influx of glutamine, phenylalanine and methionine (equals to $2$ mmol/(gDWh)) and a variable overall glucose supply. 

All objective functions we tested yield a lactate overflow with lactate shuttle towards the acceptor, suggesting that the crowding constraint is responsible for the effect.
In particular, we recover high levels of lactate shuttling for palmitate optimization and moderate levels for the optimization of the production of amino acids.
The only lactate shuttle truly related to an energetic imbalance is the one induced by palmitate optimization, while the others are similar in magnitude to the
shuttling present in the absence of any optimization (i.e. for $\beta=0$). 
In Fig.~S2
, we display the production flux  and lactate shuttling profiles as a function of the total glucose supply for the two-cell system
 when the lactate donor maximizes  palmitate, proline, cysteine, and serine, respectively. Other biosynthetic objective functions lead to similar features.

\clearpage

\widetext

\setcounter{figure}{0}
\makeatletter 
\renewcommand{\thefigure}{S\@arabic\c@figure}
\renewcommand{\thetable}{S\@arabic\c@table}
\makeatother

\section*{Supplementary Figures}

\begin{figure}[h]
\centering
\includegraphics[width=0.3\textwidth]{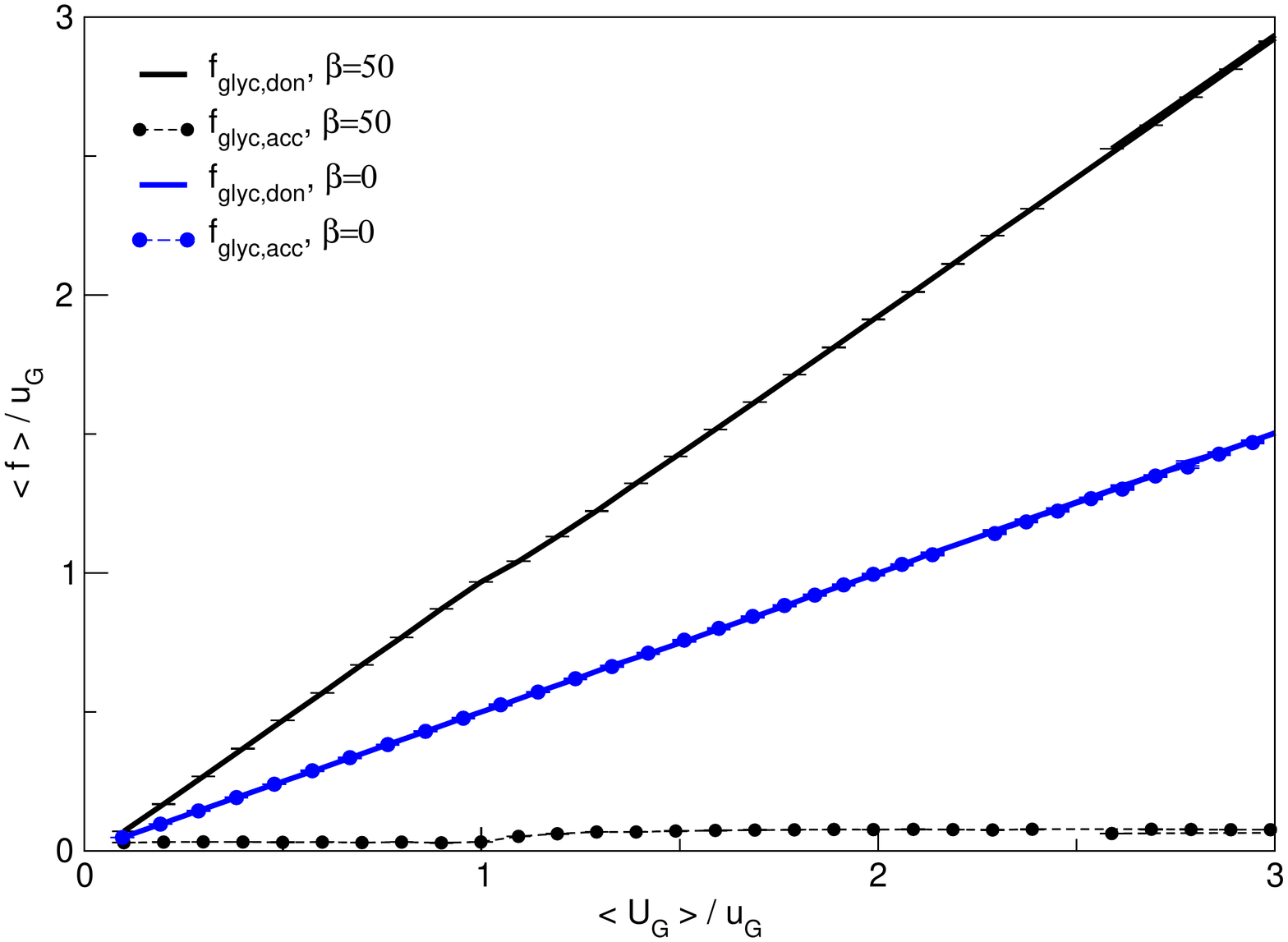}
\includegraphics[width=0.3\textwidth]{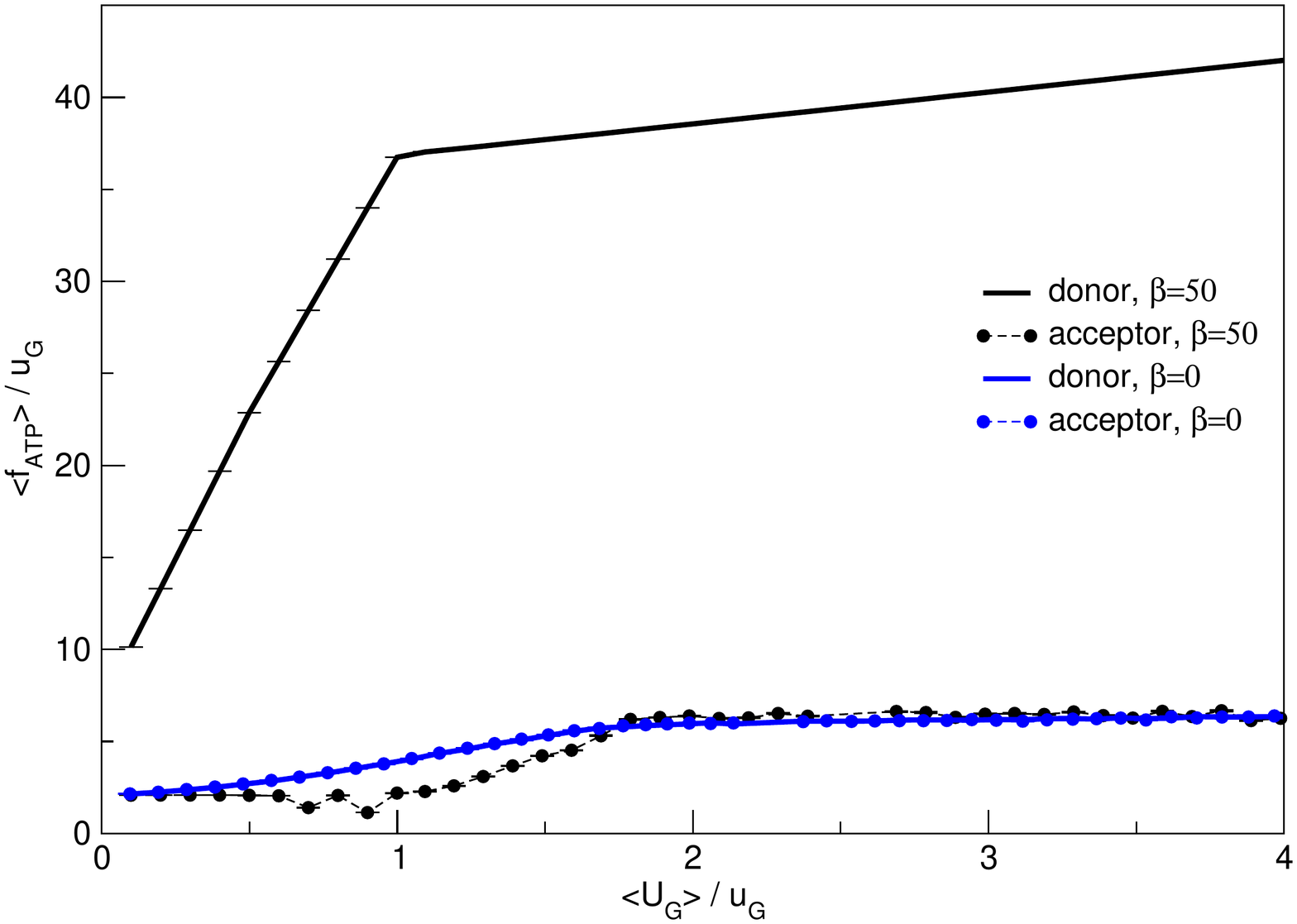}\\
\includegraphics[width=0.3\textwidth]{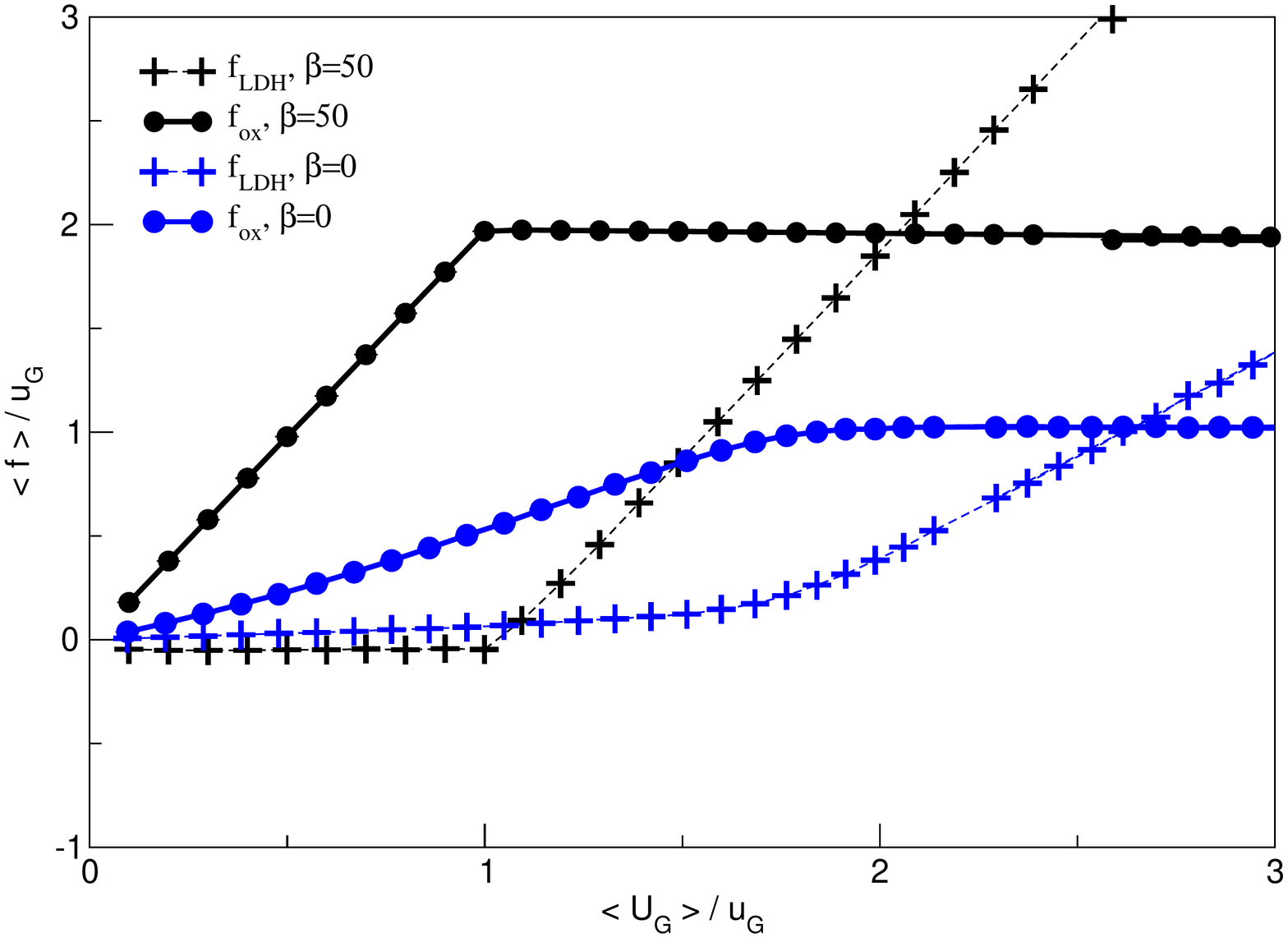}
\includegraphics[width=0.3\textwidth]{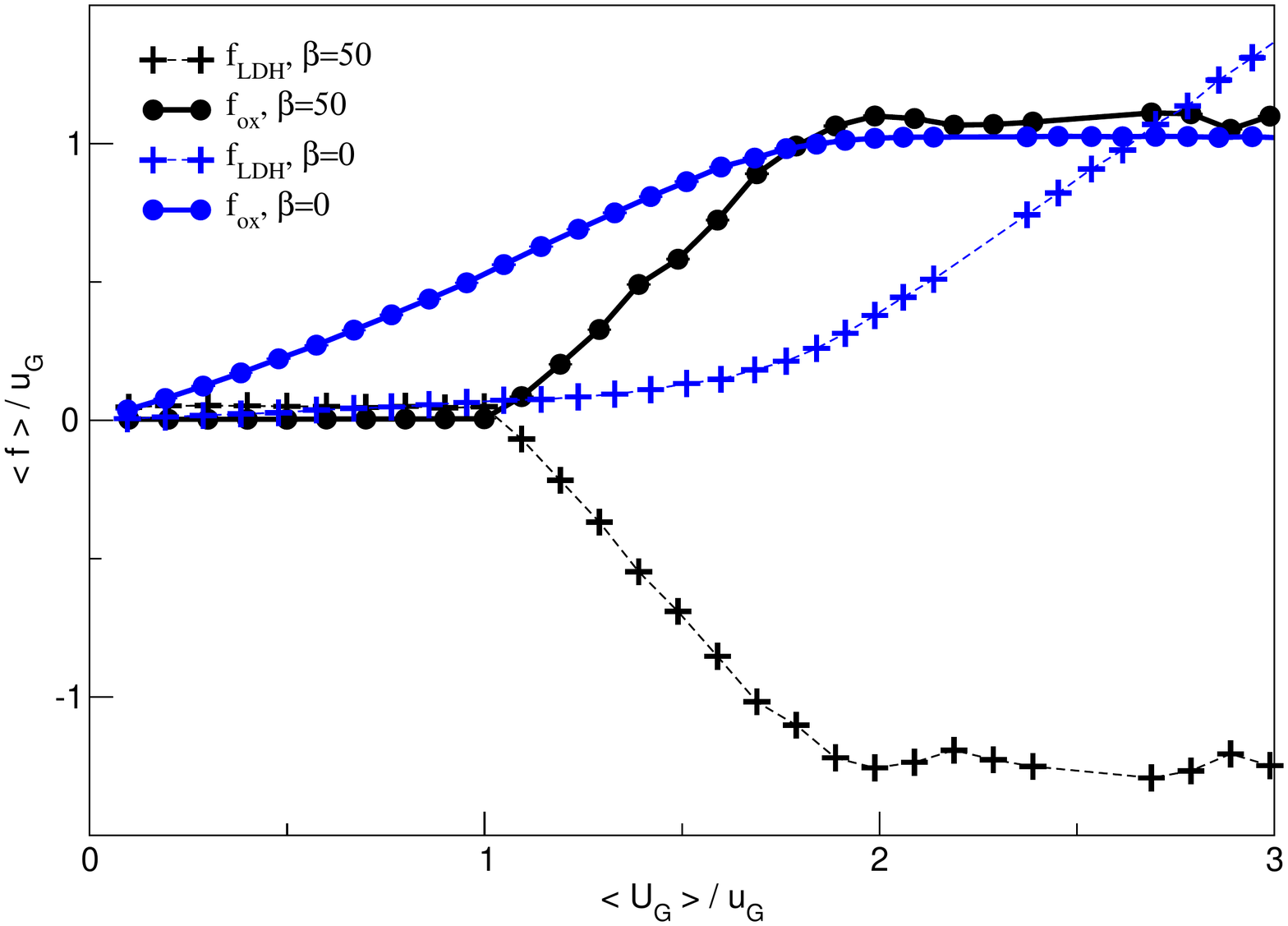}
\caption{
\textbf{Glucose intake, ATP production, and oxidative and fermentative fluxes in a donor-acceptor symmetric system.}
(a, top left) Glucose intakes for two coupled symmetric cells as a function of the total glucose available to the donor-acceptor pair.
(b, top right) ATP produced by the donor and the acceptor cells as a function of the total glucose available to the donor-acceptor pair.
(c, bottom left)--(d, bottom right) Average flux through LDH (circles) and PDHm (crosses) as a function of the average glucose supplied to the two-cell system. 
Curves describe the behaviour obtained for two coupled symmetric HCCN cells with an ATP-maximizing donor (black lines, $\beta=50$) or for an unbiased sampling of the two-cell solution space (blue lines, $\beta=0$).
Error bars, which represents s.e.m., are smaller than symbol sizes.
\label{FIG:S}
}
\end{figure}

\begin{figure}[h!!!]
\centering
\includegraphics[width=0.5\paperwidth]{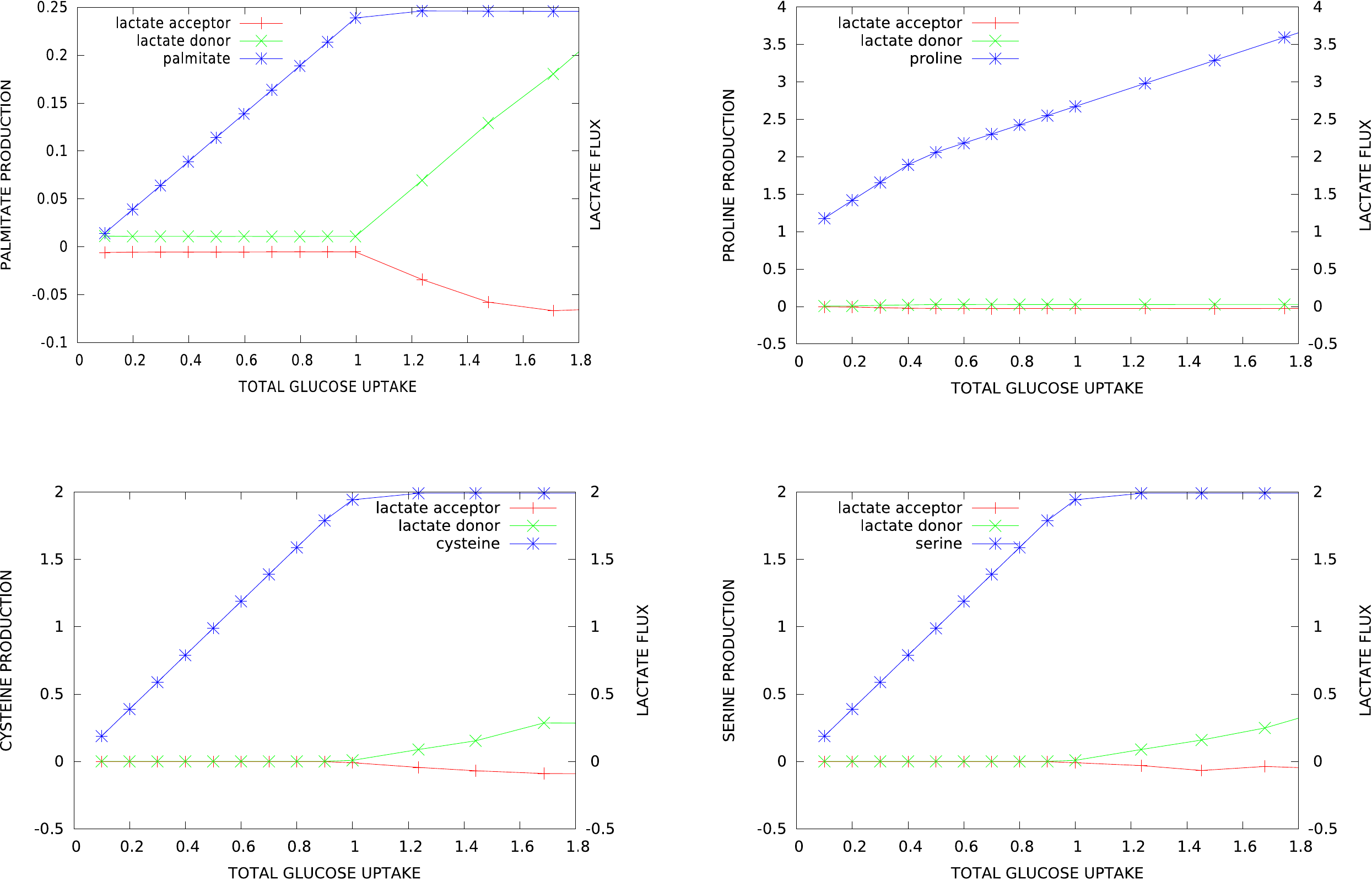}
\caption{\textbf{Lactate shuttling for alternative objective function maximizations.} In blue we plot the flux that is maximized in the donor cell: palmitate (top left), proline (top right), cysteine (bottom left), and serine (bottom right) as a function of the total glucose supply, for a system formed by a lactate donor and a lactate acceptor. The lactate fluxes of donor and acceptor cells are depicted in green and red, respectively. A negative flux correspond to lactate influx.}
\label{other_objective}
\end{figure}

\begin{figure}
\centerline{\includegraphics[width=0.5\paperwidth]{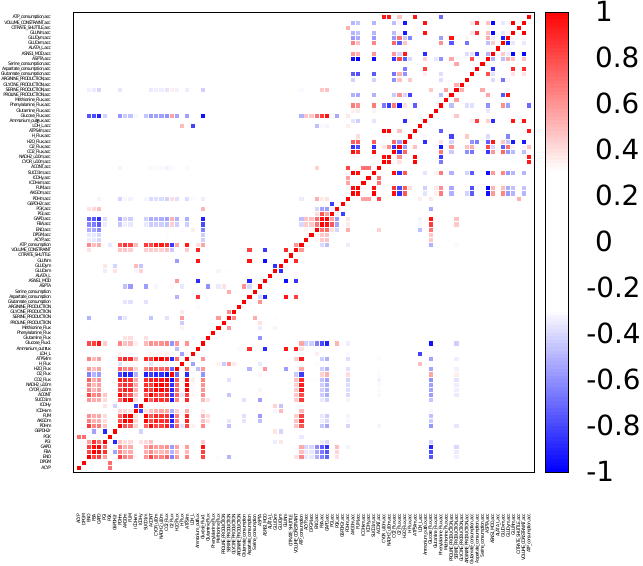}}
\caption{\textbf{Pearson correlation coefficients among all fluxes of lactate donor and lactate acceptor for maximum glucose uptake $U_G=0.3$.} The fluxes of the acceptor cell are identified by the trailing letters ``acc".}
\end{figure}

\begin{figure}
\centerline{\includegraphics[width=0.5\paperwidth]{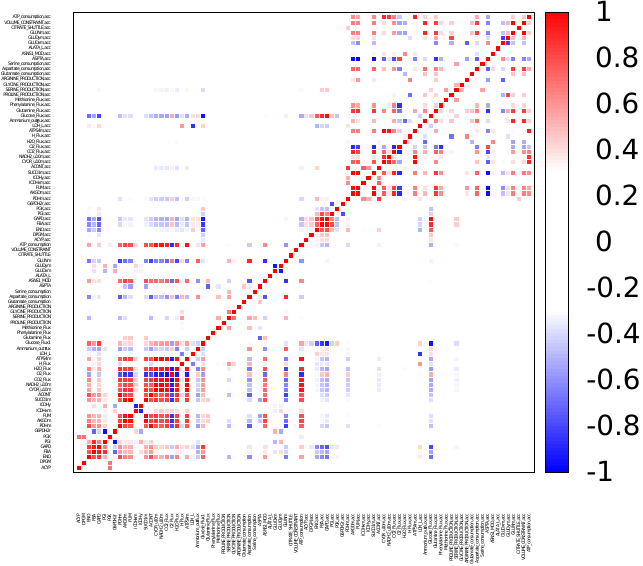}}
\caption{\textbf{Pearson correlation coefficients among all fluxes of lactate donor and lactate acceptor for maximum glucose uptake $U_G=1$.} The fluxes of the acceptor cell are identified by the trailing letters ``acc".}
\end{figure}

\begin{figure}
\centerline{\includegraphics[width=0.5\paperwidth]{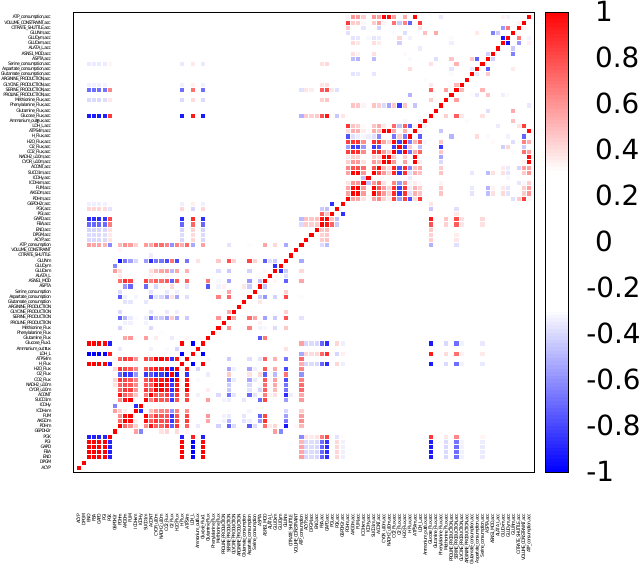}}
\caption{\textbf{Pearson correlation coefficients among all fluxes of lactate donor and lactate acceptor for maximum glucose uptake $U_G=3$.} The fluxes of the acceptor cell are identified by the trailing letters ``acc".}
\end{figure}

\begin{figure}
\centerline{\includegraphics[width=0.5\paperwidth]{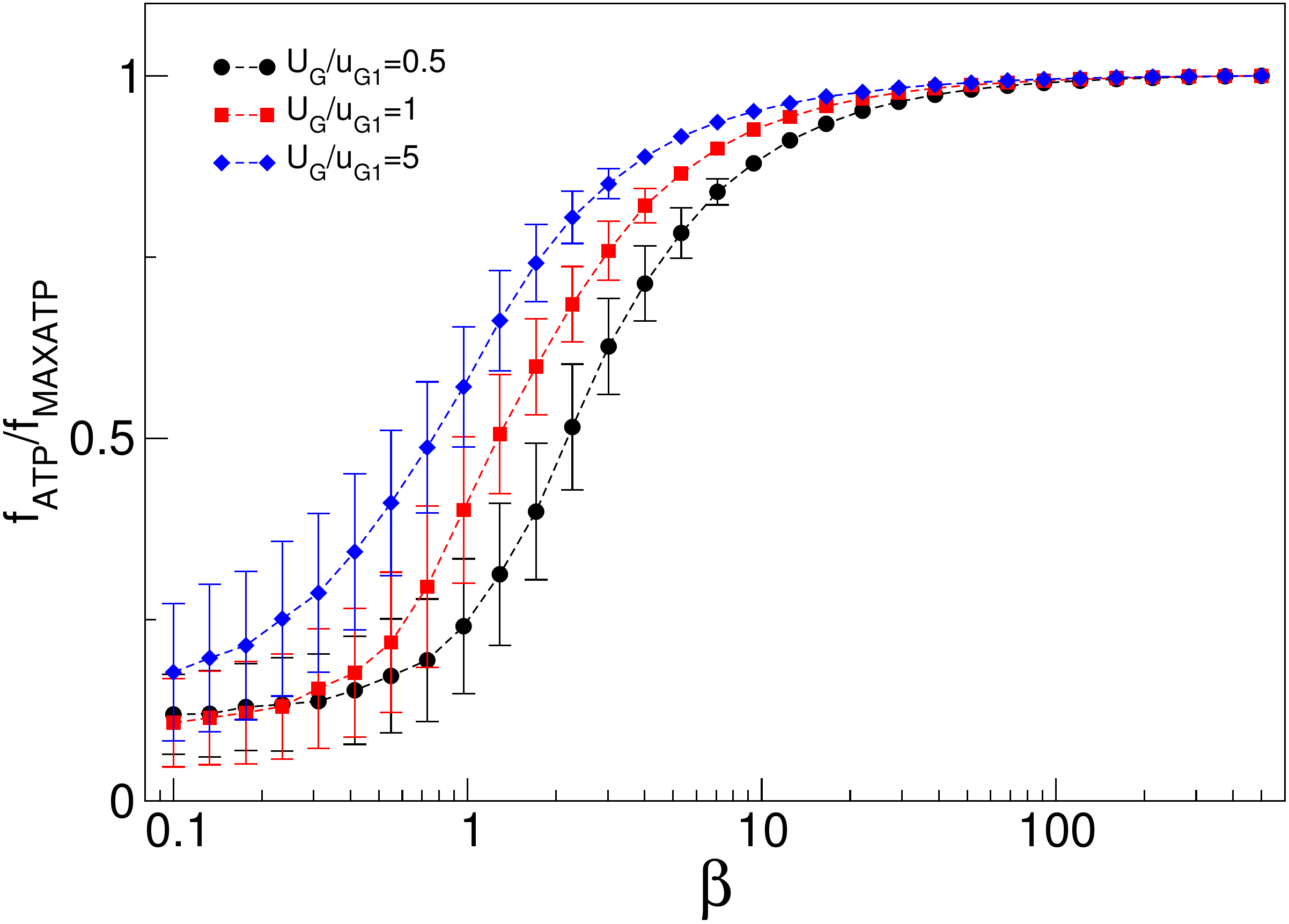}}
\caption{\textbf{By increasing $\beta$, an HCCN increases the ATP production eventually saturating the capacity at around $\beta\simeq 50$.} We display normalized ATP production fluxes for maximal glucose supply $U_{MAX}$ equals to $0.5$ (black circels), $1$ (red squares), and $5$ (blu diamonds).}
\label{fig: ATP vs BETA}
\end{figure}

\clearpage

\section*{Supplementary Tables}

\begin{table}[h!]
\centering
\scalebox{0.75}{
\begin{tabular}{| l | l |}
\hline
Variable symbol & Flux identified \\
\hline
$U_\G$   &   Glucose supply  \\
\hline
$f_\GL$, $f_\HEX$   &   Glycolysis flux  \\
\hline
$f_\M$, $f_\PDH$   &   Oxidative phosphorylation flux  \\
\hline
$f_\LDH$   &   Flux through lactate dehydrogenase  \\
\hline
$f_\ATP$   &   ATP production  \\
\hline
\end{tabular}
}
\caption{\textbf{List of more relevant variables appearing in the manuscript.}}
\label{TAB: more relevant variables}
\end{table}

\begin{table}
\centering
\scalebox{0.75}{
\begin{tabular}{| l | l |}

\hline
Metabolite short name & Metabolite extended name \\
\hline
\ce{13DPG}   &   3-Phospho-D-glyceroyl phosphate  \\
\hline
\ce{23DPG}   &   2,3-Disphospho-D-glycerate  \\
\hline
\ce{2PG}   &   D-Glycerate 2-phosphate  \\
\hline
\ce{3PG}   &   3-Phospho-D-glycerate  \\
\hline
\ce{6PGC}   &   6-Phospho-D-gluconate  \\
\hline
\ce{6PGL}   &   6-phospho-D-glucono-1,5-lactone  \\
\hline
\ce{ACCoA}   &   Acetyl-CoA  \\
\hline
\ce{ADP}   &   Adenosine diphosphate  \\
\hline
\ce{AKG}   &   2-Oxoglutarate  \\
\hline
\ce{ATP}   &   Adenosine triphosphate  \\
\hline
\ce{CIT}   &   Citrate  \\
\hline
\ce{CO2}   &   Carbon dioxyde  \\
\hline
\ce{CoA}   &   Coenzyme A  \\
\hline
\ce{DHAP}   &   Dihydroxyacetone phosphate  \\
\hline
\ce{E 4 P}   &   D-Erythrose 4-phosphate  \\
\hline
\ce{F 6 P}   &   D-Fructose 6-phosphate  \\
\hline
\ce{FAD}   &   Flavin adenine dinucleotide oxidized  \\
\hline
\ce{FADH2}   &   Flavin adenine dinucleotide reduced  \\
\hline
\ce{FDP}   &   D-Fructose 1,6-bisphosphate  \\
\hline
\ce{FICYTC}   &   Ferricytochrome C  \\
\hline
\ce{FOCYTC}   &   Ferrocytochrome C  \\
\hline
\ce{FUM}   &   Fumarate  \\
\hline
\ce{G 3 P}   &   Glyceraldehyde 3-phosphate  \\
\hline
\ce{G 6 P}   &   D-Glucose 6-phosphate  \\
\hline
\ce{GLC}   &   D-Glucose  \\
\hline
\ce{H2O}   &   Water molecule \\
\hline
\ce{H}[M]   &   Hydrogen ion as an electromotive force (mitochondrial)  \\
\hline
\ce{H}   &   Hydrogen ion in cytoplasm\\
\hline
\ce{ICIT}   &   Isocitrate  \\
\hline
\ce{LAC-L}   &   L-Lactate  \\
\hline
\ce{MAL-L}   &   L-Malate  \\
\hline
\ce{NAD}   &   Nicotinamide adenine dinucleotide  \\
\hline
\ce{NADH}   &   Nicotinamide adenine dinucleotide - reduced  \\
\hline
\ce{NADP}   &   Nicotinamide adenine dinucleotide phosphate  \\
\hline
\ce{NADPH}   &   Nicotinamide adenine dinucleotide phosphate - reduced  \\
\hline
\ce{O2}   &   Molecular oxygen  \\
\hline
\ce{O2S}   &   Superoxide anion  \\
\hline
\ce{OAA}   &   Oxaloacetate  \\
\hline
\ce{PEP}   &   Phosphoenolpyruvate  \\
\hline
\ce{Pi}   &   Phosphate  \\
\hline
\ce{PYR}   &   Pyruvate  \\
\hline
\ce{Q 10}   &   Ubiquinone-10  \\
\hline
\ce{Q 10H2}   &   Ubiquinol-10  \\
\hline
\ce{R 5 P}   &   alpha-D-Ribose 5-phosphate  \\
\hline
\ce{RU 5 P-D}   &   D-Ribulose 5-phosphate  \\
\hline
\ce{S 7 P}   &   Sedoheptulose 7-phosphate  \\
\hline
\ce{SUCC}   &   Succinate  \\
\hline
\ce{SUCCoA}   &   Succinyl-CoA  \\
\hline
\ce{XU 5 P-D}   &   D-Xylulose 5-phosphate  \\
\hline
\ce{gln\_L} & Glutamine \\
\hline
\ce{glu\_L} & Glutamate \\
\hline
\ce{NH4} & Ammonia \\
\hline
\ce{CH3} & Methil group \\
\hline 
\ce{asp\_L} & Aspartate \\
\hline
\ce{ala\_L} & Alanine \\
\hline
\ce{asn\_L} & Asparagine \\
\hline
\ce{pro\_L} & Proline \\
\hline
\ce{ser\_L} & Serine \\
\hline
\ce{gly} & Glycine \\
\hline
\ce{arg\_L} & Arginine \\
\hline
\ce{cys\_L} & Cysteine \\
\hline 
\ce{met\_L} & Methionine \\
\hline
\ce{tyr\_L} & Tyrosine \\
\hline
\ce{phe\_L} & Phenylalanine \\
\hline
\ce{hdca} & Palmitic acid \\
\hline
\end{tabular}
}
\caption{\textbf{List of metabolites appearing in a single HCCN model.}}
\label{TAB: metabolites}
\end{table}

\begin{table}
\centering %
\begin{tabular}{|l|l|}
\hline 
Enzyme  & Reaction \\
\hline 
ACONT  & \ce{ CIT  <-> ICIT } \\
\hline 
ACYP  & \ce{ 13DPG + H2O  ->  3PG + H + Pi} \\
\hline 
AKGDm  & \ce{ AKG + CoA + NAD  ->   CO2 + NADH + SUCCoA} \\
\hline 
ATPS4 m  & \ce{ADP + 4 H + Pi  ->  ATP + 3 H[M] + H2O} \\
\hline 
CSm  & \ce{ACCoA + H2O + OAA  ->   CIT + CoA + H[M]} \\
\hline 
CYOOm3  & \ce{4 focytC + $7.92$ H[M] + O2  -> 4 ficytC + 4 H + $1.96$ H2O + $0.02$ O2-} \\
\hline 
CYOR{\textunderscore}u10m  & \ce{2 ficytC + 2 H[M] + Q10H2  -> 2 focytC + 4 H + Q10} \\
\hline 
DPGM  & \ce{13DPG   <->  23DPG + H} \\
\hline 
DPGase  & \ce{23DPG + H2O  ->   3PG + Pi} \\
\hline 
ENO  & \ce{2PG  <-> H2O + PEP} \\
\hline 
FBA  & \ce{FDP  <->  DHAP + G 3 P} \\
\hline 
FUM  & \ce{FUM + H2O   <-> MAL-L} \\
\hline 
G6PDH2r  & \ce{G 6 P + NADP  <->  6PGL + H + NADPH} \\
\hline 
GAPD  & \ce{G 3 P + NAD + Pi   <->  13DPG + H + NADH} \\
\hline 
GND  & \ce{6PGC + NADP  ->   CO2 + NADPH + RU 5 P-D} \\
\hline 
HEX1  & \ce{ATP + GLC  ->   ADP + G 6 P + H} \\
\hline 
ICDHxm  & \ce{ICIT + NAD ->   AKG + CO2 + NADH} \\
\hline 
ICDHy  & \ce{ICIT + NADP  ->   AKG + CO2 + NADPH} \\
\hline 
LDH  & \ce{LAC-L + NAD   <->  H + NADH + PYR} \\
\hline 
MDH  & \ce{MAL-L + NAD   <->  H + NADH + OAA} \\
\hline 
NADH2{\textunderscore}u10m  & \ce{5 H + NADH + Q 10  ->   4 H + NAD + Q 10H2} \\
\hline 
PDHm  & \ce{CoA + NAD + PYR  -> ACCoA + CO2 + NADH} \\
\hline 
PFK  & \ce{ATP + F 6 P  -> ADP + FDP + H} \\
\hline 
PGI  & \ce{G 6 P  <-> F 6 P} \\
\hline 
PGK  & \ce{3PG + ATP <-  13DPG + ADP} \\
\hline 
PGL  & \ce{6PGL +  H2O  ->  6PGC + H} \\
\hline 
PGM  & \ce{2PG  <->  3PG} \\
\hline 
PYK  & \ce{ADP + H + PEP  -> ATP + PYR} \\
\hline 
RPE  & \ce{RU 5 P-D   <-> XU 5 P-D} \\
\hline 
RPI  & \ce{ R 5 P  <->  RU 5 P-D} \\
\hline 
SUCD1m  & \ce{FAD + SUCC  <-> FADH2 + FUM} \\
\hline 
SUCOASm  & \ce{ATP + CoA + SUCC   <->  ADP + Pi + SUCCoA} \\
\hline 
TALA  & \ce{G 3 P + S 7 P   <->  E 4 P + F 6 P} \\
\hline 
TKT1  & \ce{R 5 P + XU 5 P-D   <->  G 3 P + S 7 P} \\
\hline 
TKT2  & \ce{E 4 P + XU 5 P-D   <-> F 6 P  + G 3 P} \\
\hline 
TPI  & \ce{DHAP   <->  G 3 P} \\
\hline 
O2S reduction & \ce{NADPH + O2-  + 2H  ->  NADP + 2 H2O} \\
\hline 
FAD regeneration  & \ce{Q 10 + FADH2  ->   Q 10 H2  + FAD } \\
\hline 
ATP consumption & \ce{ATP + H2O  ->   ADP + Pi + H} \\
\hline 
\end{tabular}\protect\caption{\textbf{Internal reactions of an HCCN with the corresponding enzyme that catalyzes it: main catabolic pathways.}}
\label{TAB: internal reactions one}
\end{table}

\begin{table}
\centering %
\begin{tabular}{|l|l|}
\hline 
Enzyme  & Reaction \\
\hline 
ALATA\_L  & \ce{ AKG + ala\_L <-> glu\_L + PYR} \\
\hline 
ASPTA  & \ce{AKG + asp\_L <-> glu\_L + OAA } \\
\hline 
GLUDxm  & \ce{glu\_L + H2O + NAD -> AKG + H + NADH + NH4 } \\
\hline 
GLUDym  & \ce{glu\_L + H2O + NADP -> AKG + H + NADPH + NH4} \\
\hline   
GLUN  & \ce{gln\_L + H2O -> glu\_L + NH4} \\
\hline 
Tyrosine production & \ce{ phe\_L + NADH + O2 + h  ->  tyr\_L + NAD + H2O} \\
\hline
Asparagine production & \ce{asp\_L + (2) ATP + gln\_L + (2) H2O -> (2) ADP + asn\_L + glu\_L + (2) H + (2) Pi } \\
\hline
Proline production & \ce{glu\_L + ATP + (2) NADPH + 2 H^+ ->  pro\_L + ADP + Pi + H2O + (2) NADP } \\
\hline
Serine production & \ce{3PG + NAD + H2O + glu\_L ->  ser\_L + AKG + NADH + Pi + H } \\
\hline
Glycine production & \ce{ ser\_L + H2O + (2) NADP   -> gly + CO2 + (2) NADPH + (2) H} \\
\hline
Arginine production &\begin{tabular}[c]{@{}c@{}}
\ce{asp\_L + glu\_L + gln\_L + (5) ATP + NADPH +  3 H2O + CO2 ->} \\
\ce{arg\_L + FUM + AKG +  (5)ADP + (5) Pi + 5 H + NADP } \\
\end{tabular}\\   
\hline
Cysteine production &\begin{tabular}[c]{@{}c@{}}
\ce{met\_L + ser\_L + COA + (4) ATP + NAD + CO2 + (4) H2O ->} \\
\ce{ cys\_L + SUCCOA + (4)ADP + (4)Pi + 4 H + NADH + NH4 + CO2 + CH3  } \\
\end{tabular}\\
\hline
Palmitic acid production &\begin{tabular}[c]{@{}c@{}}
\ce{(8)ACCOA + (23) ATP + 8 NADH + (6) NADPH + 17 H2O ->} \\
\ce{ hdca + (8)COA + (8)NAD + (6) NADP + (23) ADP  + (23) Pi  + (10) H } \\
\end{tabular}\\
\hline 
\end{tabular}\caption{\textbf{Internal reactions of an HCCN with the corresponding enzyme that catalyzes it: glutamate metabolism and anabolic effective reactions.}}
\label{TAB: internal reactions}
\end{table}

\begin{table}
\centering %
\begin{tabular}{|l|c|c|l|}
\hline 
Description  & Lower bound & Upper bound & Reaction \\
\hline 
Flux of carbon dioxide & -1000 & 1000 & $\mathrm{CO_{2}}$$\longleftrightarrow$ out \\
\hline 
Flux of oxygen & -1000 & 1000 & $\mathrm{O_{2}}$ $\longleftrightarrow$ out \\
\hline 
Flux of water  & -1000 & 1000 & $\mathrm{H_{2}O}$ $\longleftrightarrow$ out \\
\hline 
Flux of ion hydrogen & -1000 & 1000 & $\mathrm{H}$ $\longleftrightarrow$ out \\
\hline 
Glucose intake  & 0 & $\mbox{\ensuremath{U_{G}}}$ & in $\longleftarrow$ GLC \\
\hline 
Lactate flux  & %
\begin{tabular}[c]{@{}c@{}}
0\\
-1000 \\
\end{tabular} & %
\begin{tabular}[c]{@{}c@{}}
1000\\
1000 \\
\end{tabular} & %
\begin{tabular}[c]{@{}c@{}}
LAC-L $\longrightarrow$ out (donor) \\
LAC-L $\longleftrightarrow$ out (acceptor) \\
\end{tabular}\\
\hline 
Glutamine intake  & 0 & 1 & in $\longleftarrow$ gln\_L \\
\hline
Phenylalanine intake  & 0 & 1000 &in $\longleftarrow$ phe\_L \\
\hline
Methionine intake  & 0 & 1000 & in $\longleftarrow$ met\_L \\
\hline
Ammonia flux  & -1000 & 1000 & \ce{NH4} $\leftrightarrow$ out \\
\hline
Methyl group flux  & -1000 & 1000 & \ce{CH3} $\leftrightarrow$ out\\
\hline
\end{tabular}
\caption{\textbf{Exchange reactions of an HCCN with the corresponding bounds.}
In this model, we assume that carbon dioxide, water, oxygen, and proton can freely diffuse in and out of the cell. The maximum glucose intake for each cell cannot exceed the total glucose supply.
Lactate flux distinguishes donor and acceptor cells as the donor can only secrete lactate, while the acceptor can also intake lactate.
 In Table \ref{TAB: extra constraints}, the constraints on glucose and lactate fluxes for the donor-acceptor couple are presented.
\label{TAB: exchange reactions}}
\end{table}

\begin{table}
\centering %
\begin{tabular}{|l|c|c|l|}
\hline 
Enzyme & Lower bound & Upper bound & Enzyme extended name\\
\hline 
ACONT  & -1000 & 1000 & Aconitase \\
\hline 
ACYP  & 0 & 1000 & Acylphosphatase \\
\hline 
AKGDm  & 0 & 1000 & 2-Oxoglutarate dehydrogenase \\
\hline 
ATPS4m  & 0 & 1000 & ATP synthase \\
\hline 
CSm  & 0 & 1000 & Citrate synthase \\
\hline 
CYOOm3  & 0 & 1000 & Cytochrome C oxidase, mitochondrial Complex IV \\
\hline
CYOR{\textunderscore}u10m  & 0 & 1000 & Ubiquinol-6 cytochrome C reductase, Complex III \\
\hline
DPGM  & -1000 & 1000 & Diphosphoglyceromutase \\
\hline 
DPGase  & 0 & 1000 & Diphosphoglycerate phosphatase \\
\hline 
ENO  & -1000 & 1000 & Enolase \\
\hline 
FBA  & -1000 & 1000 & Fructose-bisphosphate aldolase \\
\hline 
FUM  & -1000 & 1000 & Fumarase \\
\hline 
G6PDH2r  & -1000 & 1000 & Glucose 6-phosphate dehydrogenase \\
\hline 
GAPD  & -1000 & 1000 & Glyceraldehyde-3-phosphate dehydrogenase \\
\hline 
GND  & 0 & 1000 & Phosphogluconate dehydrogenase \\
\hline 
HEX1  & 0 & 1000 & Hexokinase \\
\hline 
ICDHxm  & 0 & 1000 & Isocitrate dehydrogenase \\
\hline 
ICDHy  & 0 & 1000 & Isocitrate dehydrogenase \\
\hline 
LDH  & -1000 & 1000 & Lactate dehydrogenase \\
\hline 
MDH  & -1000 & 1000 & Malate dehydrogenase \\
\hline 
NADH2{\textunderscore}u10m  & 0 & 1000 & NADH dehydrogenase, mitochondrial \\
\hline 
PDHm  & 0 & 1000 & Pyruvate dehydrogenase \\
\hline 
PFK  & 0 & 1000 & Phosphofructokinase \\
\hline 
PGI  & -1000 & 1000 & Glucose-6-phosphate isomerase \\
\hline 
PGK  & -1000 & 0 & Phosphoglycerate kinase \\
\hline 
PGL  & 0 & 1000 & 6-phosphogluconolactonase \\
\hline 
PGM  & -1000 & 1000 & Phosphoglycerate mutase \\
\hline 
PYK  & 0 & 1000 & Pyruvate kinase \\
\hline 
RPE  & -1000 & 1000 & Ribulose-5-phosphate 3-epimerase \\
\hline 
RPI  & -1000 & 1000 & Ribose-5-phosphate isomerase \\
\hline 
SUCD1m  & -1000 & 1000 & Succinate dehydrogenase \\
\hline 
SUCOASm  & -1000 & 1000 & Succinate--CoA ligase \\
\hline 
TALA  & -1000 & 1000 & Transaldolase \\
\hline 
TKT1  & -1000 & 1000 & Transketolase \\
\hline 
TKT2  & -1000 & 1000 & transketolase \\
\hline 
TPI  & -1000 & 1000 & Triose-phosphate isomerase \\
\hline 
Lumped reaction & 0 & 1000 & Reduction of superoxyde anion \\
\hline 
Lumped reaction  & 0 & 1000 & FAD regeneration \\
\hline 
Effective reaction & $\atpmin$ & 1000 & ATP hydrolysis \\
\hline 
\end{tabular}
\caption{\textbf{Internal reactions of an HCCN with the corresponding bounds: catabolic pathways.}
\label{TAB: bounds one}}
\end{table}

\begin{table}
\centering %
\begin{tabular}{|l|c|c|l|}
\hline 
Enzyme & Lower bound & Upper bound & Enzyme extended name\\
\hline 
ALATA\_L  & -1000 &  1000 & alanine transaminase \\
\hline 
ASPTA  & -1000 &  1000 & aspartate transaminase \\
\hline 
GLUDxm  & -1000 &  1000 & glutamate dehydrogenase (NAD) \\
\hline 
GLUDym  & 0 &  1000 & glutamate dehydrogenase (NADP) \\
\hline  
GLUN   & 0 &  1000 & glutaminase \\
\hline 
Lumped reaction & 0 &  1000 & Tyrosine production \\
\hline
 Lumped reaction& 0 &  1000 & Asparagine production \\
\hline
Lumped reaction & 0 &  1000 & Proline production \\
\hline
Lumped reaction & 0 &  1000 & Serine production \\
\hline
Lumped reaction & 0 &  1000 & Glycine production \\
\hline
Lumped reaction & 0 &  1000 & Arginine production \\
\hline
Lumped reaction & 0 &  1000 & Cysteine production \\
\hline
Lumped reaction & 0 &  1000 & Palmitic acid production \\
\hline 
\end{tabular}\protect\caption{\textbf{Internal reactions of an HCCN with the corresponding bounds: glutamate metabolism and anabolic effective reactions.}
\label{TAB: bounds}}
\end{table}

\begin{table}
\centering %
\begin{tabular}{|l|c|c|l|}
\hline
Variable constrained  & Lower bound & Upper bound & Variable expressed by reactions \\
\hline
Donor resource & 0 & $\Phi_{\ATP}$  & $a_{\GL}f_{\HEX,\D}+a_{\M}f_{\PDH,\D}+a_{\L}f_{\LDH,\D}$ \\
\hline
Acceptor resource & 0 & $\Phi_{\ATP}$ & $a_{\GL}f_{\HEX,\A}+a_{\M}f_{\PDH,\A}+a_{\L}|f_{\LDH,\A}|$ \\
\hline
Total glucose intake & 0 & $U_{\G}$ & $U_{\G,\D}+U_{\G,\A}$ \\
\hline
Total lactate flux & 0 & 1000 & $U_{\LAC,\D}+U_{\LAC,\A}$ \\
\hline
\end{tabular}
\caption{\textbf{Constraints on maximum resources available for ATP production and on total glucose and lactate fluxes.}
The first column contains a description of the variable that is constrained,
while the last column the variable expressed as a function of the elementary fluxes of the metabolic network.
The second and the third column contain the minimum and the maximum value that the variable can take, respectively.
$\Phi_{\ATP}$ is set to 0.4 and states that at most 40\% of the cellular resources can be devoted to ATP production.
The sum of the glucose intaken by donor and acceptor cannot exceed the total glucose supply $U_{\G}$. 
The constraint on total lactate flux establishes that there is no external source of lactate in the system and that the donor-acceptor couple can only produce lactate.}
\label{TAB: extra constraints}
\end{table}


\begin{table*}
\centering
\begin{tabular}{  | l | l  |}
\hline
Metabolite & Biomass coefficient\\
\hline
\ce{H2O} &  -20.651 \\
\ce{ATP} &  -20.651 \\
\ce{ADP} &  20.651 \\
\ce{H} & 20.651 \\
\ce{Pi} & 20.971 \\
glu\_L &  -0.38587	\\
asp\_L &  -0.35261	\\
asn\_L &  -0.27942	\\
ala\_L & -0.50563	\\
cys\_L & -0.046571	\\
gln\_L &  -0.326	\\
gly & -0.53889	\\
ser\_L & -0.39253	\\
arg\_L & -0.35926	\\
met\_L & -0.15302	\\
tyr\_L & -0.15967	\\
phe\_L & -0.25947	\\
pro\_L & -0.41248	\\
hdca	 & -0.112		\\
\ce{R 5 P} & -0.045 \\
\ce{G 6 P} & -0.275 \\
\hline
\end{tabular}
\caption{\textbf{Biomass objective function coefficients adapted from [47]. }}
\label{TableShortNames}
\end{table*}


\begin{thebibliography}{10}
\expandafter\ifx\csname url\endcsname\relax
  \def\url#1{\texttt{#1}}\fi
\expandafter\ifx\csname urlprefix\endcsname\relax\def\urlprefix{URL }\fi
\providecommand{\bibinfo}[2]{#2}
\providecommand{\eprint}[2][]{\url{#2}}

\bibitem{Molenaar:2009p3978}
\bibinfo{author}{Molenaar, D.}, \bibinfo{author}{{van Berlo}, R.},
  \bibinfo{author}{{de Ridder}, D.} \& \bibinfo{author}{Teusink, B.}
\newblock \bibinfo{title}{Shifts in growth strategies reflect tradeoffs in
  cellular economics}.
\newblock \emph{\bibinfo{journal}{Mol Syst Biol}} \textbf{\bibinfo{volume}{5}},
  \bibinfo{pages}{323} (\bibinfo{year}{2009}).

\bibitem{Paczia:2012p4266}
\bibinfo{author}{Paczia, N.} \emph{et~al.}
\newblock \bibinfo{title}{Extensive exometabolome analysis reveals extended
  overflow metabolism in various microorganisms}.
\newblock \emph{\bibinfo{journal}{Microb Cell Fact}}
  \textbf{\bibinfo{volume}{11}}, \bibinfo{pages}{122} (\bibinfo{year}{2012}).

\bibitem{Hsu:2008p4267}
\bibinfo{author}{Hsu, P.~P.} \& \bibinfo{author}{Sabatini, D.~M.}
\newblock \bibinfo{title}{Cancer cell metabolism: {Warburg} and beyond}.
\newblock \emph{\bibinfo{journal}{Cell}} \textbf{\bibinfo{volume}{134}},
  \bibinfo{pages}{703--7} (\bibinfo{year}{2008}).

\bibitem{Kroemer:2008p3885}
\bibinfo{author}{Kroemer, G.} \& \bibinfo{author}{Pouyssegur, J.}
\newblock \bibinfo{title}{Tumor cell metabolism: cancer's achilles' heel}.
\newblock \emph{\bibinfo{journal}{Cancer Cell}} \textbf{\bibinfo{volume}{13}},
  \bibinfo{pages}{472--82} (\bibinfo{year}{2008}).

\bibitem{Funes:2007p4268}
\bibinfo{author}{Funes, J.~M.} \emph{et~al.}
\newblock \bibinfo{title}{Transformation of human mesenchymal stem cells
  increases their dependency on oxidative phosphorylation for energy
  production}.
\newblock \emph{\bibinfo{journal}{Proc Natl Acad Sci USA}}
  \textbf{\bibinfo{volume}{104}}, \bibinfo{pages}{6223--8}
  (\bibinfo{year}{2007}).

\bibitem{Zhou:2007p4269}
\bibinfo{author}{Zhou, S.} \emph{et~al.}
\newblock \bibinfo{title}{Frequency and phenotypic implications of
  mitochondrial {DNA} mutations in human squamous cell cancers of the head and
  neck}.
\newblock \emph{\bibinfo{journal}{Proc Natl Acad Sci USA}}
  \textbf{\bibinfo{volume}{104}}, \bibinfo{pages}{7540--5}
  (\bibinfo{year}{2007}).

\bibitem{Pouyssegur:2006p4277}
\bibinfo{author}{Pouyss{\'e}gur, J.}, \bibinfo{author}{Dayan, F.} \&
  \bibinfo{author}{Mazure, N.~M.}
\newblock \bibinfo{title}{Hypoxia signalling in cancer and approaches to
  enforce tumour regression}.
\newblock \emph{\bibinfo{journal}{Nature}} \textbf{\bibinfo{volume}{441}},
  \bibinfo{pages}{437--43} (\bibinfo{year}{2006}).

\bibitem{Christofk:2008p3930}
\bibinfo{author}{Christofk, H.} \emph{et~al.}
\newblock \bibinfo{title}{The {M2} splice isoform of pyruvate kinase is
  important for cancer metabolism and tumour growth}.
\newblock \emph{\bibinfo{journal}{Nature}} \textbf{\bibinfo{volume}{452}},
  \bibinfo{pages}{230--233} (\bibinfo{year}{2008}).

\bibitem{Levine:2010p4275}
\bibinfo{author}{Levine, A.~J.} \& \bibinfo{author}{Puzio-Kuter, A.~M.}
\newblock \bibinfo{title}{The control of the metabolic switch in cancers by
  oncogenes and tumor suppressor genes}.
\newblock \emph{\bibinfo{journal}{Science}} \textbf{\bibinfo{volume}{330}},
  \bibinfo{pages}{1340--4} (\bibinfo{year}{2010}).

\bibitem{Tennant:2010p3811}
\bibinfo{author}{Tennant, D.~A.}, \bibinfo{author}{Dur{\'a}n, R.~V.} \&
  \bibinfo{author}{Gottlieb, E.}
\newblock \bibinfo{title}{Targeting metabolic transformation for cancer
  therapy}.
\newblock \emph{\bibinfo{journal}{Nat Rev Cancer}}
  \textbf{\bibinfo{volume}{10}}, \bibinfo{pages}{267--77}
  (\bibinfo{year}{2010}).

\bibitem{VanderHeiden:2009p3807}
\bibinfo{author}{{Vander Heiden}, M.~G.}, \bibinfo{author}{Cantley, L.~C.} \&
  \bibinfo{author}{Thompson, C.~B.}
\newblock \bibinfo{title}{Understanding the {Warburg} effect: The metabolic
  requirements of cell proliferation}.
\newblock \emph{\bibinfo{journal}{Science}} \textbf{\bibinfo{volume}{324}},
  \bibinfo{pages}{1029--1033} (\bibinfo{year}{2009}).

\bibitem{Cairns:2011p3989}
\bibinfo{author}{Cairns, R.~A.}, \bibinfo{author}{Harris, I.~S.} \&
  \bibinfo{author}{Mak, T.~W.}
\newblock \bibinfo{title}{Regulation of cancer cell metabolism}.
\newblock \emph{\bibinfo{journal}{Nat Rev Cancer}}
  \textbf{\bibinfo{volume}{11}}, \bibinfo{pages}{85--95}
  (\bibinfo{year}{2011}).

\bibitem{Dang:2012p4276}
\bibinfo{author}{Dang, C.~V.}
\newblock \bibinfo{title}{Links between metabolism and cancer}.
\newblock \emph{\bibinfo{journal}{Genes Dev}} \textbf{\bibinfo{volume}{26}},
  \bibinfo{pages}{877--90} (\bibinfo{year}{2012}).

\bibitem{Vazquez:2010p3868}
\bibinfo{author}{Vazquez, A.}, \bibinfo{author}{Liu, J.},
  \bibinfo{author}{Zhou, Y.} \& \bibinfo{author}{Oltvai, Z.~N.}
\newblock \bibinfo{title}{Catabolic efficiency of aerobic glycolysis: the
  {Warburg} effect revisited}.
\newblock \emph{\bibinfo{journal}{BMC Syst Biol}} \textbf{\bibinfo{volume}{4}},
  \bibinfo{pages}{58} (\bibinfo{year}{2010}).

\bibitem{Famili:2003p4281}
\bibinfo{author}{Famili, I.}, \bibinfo{author}{Forster, J.},
  \bibinfo{author}{Nielsen, J.} \& \bibinfo{author}{Palsson, B.~O.}
\newblock \bibinfo{title}{Saccharomyces cerevisiae phenotypes can be predicted
  by using constraint-based analysis of a genome-scale reconstructed metabolic
  network}.
\newblock \emph{\bibinfo{journal}{Proc Natl Acad Sci USA}}
  \textbf{\bibinfo{volume}{100}}, \bibinfo{pages}{13134--9}
  (\bibinfo{year}{2003}).

\bibitem{Shlomi:2011p3812}
\bibinfo{author}{Shlomi, T.}, \bibinfo{author}{Benyamini, T.},
  \bibinfo{author}{Gottlieb, E.}, \bibinfo{author}{Sharan, R.} \&
  \bibinfo{author}{Ruppin, E.}
\newblock \bibinfo{title}{Genome-scale metabolic modeling elucidates the role
  of proliferative adaptation in causing the {Warburg} effect}.
\newblock \emph{\bibinfo{journal}{PLoS Comput Biol}}
  \textbf{\bibinfo{volume}{7}}, \bibinfo{pages}{e1002018}
  (\bibinfo{year}{2011}).

\bibitem{Locasale:2009p3927}
\bibinfo{author}{Locasale, J.~W.}, \bibinfo{author}{Cantley, L.~C.} \&
  \bibinfo{author}{{Vander Heiden}, M.~G.}
\newblock \bibinfo{title}{Cancer's insatiable appetite}.
\newblock \emph{\bibinfo{journal}{Nat Biotechnol}}
  \textbf{\bibinfo{volume}{27}}, \bibinfo{pages}{916--7}
  (\bibinfo{year}{2009}).

\bibitem{Deberardinis:2012p4007}
\bibinfo{author}{Deberardinis, R.~J.}
\newblock \bibinfo{title}{Good neighbours in the tumour stroma reduce oxidative
  stress}.
\newblock \emph{\bibinfo{journal}{Nat Cell Biol}}
  \textbf{\bibinfo{volume}{14}}, \bibinfo{pages}{235--6}
  (\bibinfo{year}{2012}).

\bibitem{pmid16423989}
\bibinfo{author}{Koukourakis, M.~I.}, \bibinfo{author}{Giatromanolaki, A.},
  \bibinfo{author}{Harris, A.~L.} \& \bibinfo{author}{Sivridis, E.}
\newblock \bibinfo{title}{{{C}omparison of metabolic pathways between cancer
  cells and stromal cells in colorectal carcinomas: a metabolic survival role
  for tumor-associated stroma}}.
\newblock \emph{\bibinfo{journal}{Cancer Res.}} \textbf{\bibinfo{volume}{66}},
  \bibinfo{pages}{632--637} (\bibinfo{year}{2006}).

\bibitem{Pavlides:2009p3926}
\bibinfo{author}{Pavlides, S.} \emph{et~al.}
\newblock \bibinfo{title}{The reverse {W}arburg effect: aerobic glycolysis in
  cancer associated fibroblasts and the tumor stroma}.
\newblock \emph{\bibinfo{journal}{Cell Cycle}} \textbf{\bibinfo{volume}{8}},
  \bibinfo{pages}{3984--4001} (\bibinfo{year}{2009}).

\bibitem{Duarte:2007p3938}
\bibinfo{author}{Duarte, N.~C.} \emph{et~al.}
\newblock \bibinfo{title}{Global reconstruction of the human metabolic network
  based on genomic and bibliomic data}.
\newblock \emph{\bibinfo{journal}{Proc Natl Acad Sci USA}}
  \textbf{\bibinfo{volume}{104}}, \bibinfo{pages}{1777--82}
  (\bibinfo{year}{2007}).

\bibitem{Brooks:2009p4313}
\bibinfo{author}{Brooks, G.~A.}
\newblock \bibinfo{title}{Cell-cell and intracellular lactate shuttles}.
\newblock \emph{\bibinfo{journal}{J Physiol}} \textbf{\bibinfo{volume}{587}},
  \bibinfo{pages}{5591--600} (\bibinfo{year}{2009}).

\bibitem{Barros:2013p4310}
\bibinfo{author}{Barros, L.~F.}
\newblock \bibinfo{title}{Metabolic signaling by lactate in the brain}.
\newblock \emph{\bibinfo{journal}{Trends Neurosci}}
  \textbf{\bibinfo{volume}{36}}, \bibinfo{pages}{396--404}
  (\bibinfo{year}{2013}).

\bibitem{Belanger:2011p4279}
\bibinfo{author}{B{\'e}langer, M.}, \bibinfo{author}{Allaman, I.} \&
  \bibinfo{author}{Magistretti, P.~J.}
\newblock \bibinfo{title}{Brain energy metabolism: focus on astrocyte-neuron
  metabolic cooperation}.
\newblock \emph{\bibinfo{journal}{Cell metab}} \textbf{\bibinfo{volume}{14}},
  \bibinfo{pages}{724--38} (\bibinfo{year}{2011}).

\bibitem{Dienel:2012p4280}
\bibinfo{author}{Dienel, G.~A.}
\newblock \bibinfo{title}{Brain lactate metabolism: the discoveries and the
  controversies}.
\newblock \emph{\bibinfo{journal}{J Cereb Blood Flow Metab}}
  \textbf{\bibinfo{volume}{32}}, \bibinfo{pages}{1107--38}
  (\bibinfo{year}{2012}).

\bibitem{Massucci:2013p4278}
\bibinfo{author}{Massucci, F.~A.} \emph{et~al.}
\newblock \bibinfo{title}{Energy metabolism and glutamate-glutamine cycle in
  the brain: a stoichiometric modeling perspective}.
\newblock \emph{\bibinfo{journal}{BMC Syst Biol}} \textbf{\bibinfo{volume}{7}},
  \bibinfo{pages}{103} (\bibinfo{year}{2013}).

\bibitem{pmid11737888}
\bibinfo{author}{Schor, S.~L.} \& \bibinfo{author}{Schor, A.~M.}
\newblock \bibinfo{title}{{{P}henotypic and genetic alterations in mammary
  stroma: implications for tumour progression}}.
\newblock \emph{\bibinfo{journal}{Breast Cancer Res.}}
  \textbf{\bibinfo{volume}{3}}, \bibinfo{pages}{373--379}
  (\bibinfo{year}{2001}).

\bibitem{pmid25079331}
\bibinfo{author}{Marusyk, A.} \emph{et~al.}
\newblock \bibinfo{title}{{{N}on-cell-autonomous driving of tumour growth
  supports sub-clonal heterogeneity}}.
\newblock \emph{\bibinfo{journal}{Nature}} \textbf{\bibinfo{volume}{514}},
  \bibinfo{pages}{54--58} (\bibinfo{year}{2014}).

\bibitem{Zhang:2012p4006}
\bibinfo{author}{Zhang, W.} \emph{et~al.}
\newblock \bibinfo{title}{Stromal control of cystine metabolism promotes cancer
  cell survival in chronic lymphocytic leukaemia}.
\newblock \emph{\bibinfo{journal}{Nat Cell Biol}}
  \textbf{\bibinfo{volume}{14}}, \bibinfo{pages}{276--86}
  (\bibinfo{year}{2012}).

\bibitem{Choi:2013p4423}
\bibinfo{author}{Choi, J.}, \bibinfo{author}{Kim, D.~H.},
  \bibinfo{author}{Jung, W.~H.} \& \bibinfo{author}{Koo, J.~S.}
\newblock \bibinfo{title}{Metabolic interaction between cancer cells and
  stromal cells according to breast cancer molecular subtype}.
\newblock \emph{\bibinfo{journal}{Breast Cancer Res}}
  \textbf{\bibinfo{volume}{15}}, \bibinfo{pages}{R78} (\bibinfo{year}{2013}).

\bibitem{pmid25259922}
\bibinfo{author}{Sherman, M.~H.} \emph{et~al.}
\newblock \bibinfo{title}{{{V}itamin d receptor-mediated stromal reprogramming
  suppresses pancreatitis and enhances pancreatic cancer therapy}}.
\newblock \emph{\bibinfo{journal}{Cell}} \textbf{\bibinfo{volume}{159}},
  \bibinfo{pages}{80--93} (\bibinfo{year}{2014}).

\bibitem{Draoui:2011p3948}
\bibinfo{author}{Draoui, N.} \& \bibinfo{author}{Feron, O.}
\newblock \bibinfo{title}{Lactate shuttles at a glance: from physiological
  paradigms to anti-cancer treatments}.
\newblock \emph{\bibinfo{journal}{Dis Model Mech}}
  \textbf{\bibinfo{volume}{4}}, \bibinfo{pages}{727--32}
  (\bibinfo{year}{2011}).

\bibitem{Doherty:2013p4183}
\bibinfo{author}{Doherty, J.~R.} \& \bibinfo{author}{Cleveland, J.~L.}
\newblock \bibinfo{title}{Targeting lactate metabolism for cancer
  therapeutics}.
\newblock \emph{\bibinfo{journal}{J Clin Invest}}
  \textbf{\bibinfo{volume}{123}}, \bibinfo{pages}{3685--92}
  (\bibinfo{year}{2013}).

\bibitem{pmid15736311}
\bibinfo{author}{Koukourakis, M.~I.}, \bibinfo{author}{Giatromanolaki, A.},
  \bibinfo{author}{Sivridis, E.}, \bibinfo{author}{Gatter, K.~C.} \&
  \bibinfo{author}{Harris, A.~L.}
\newblock \bibinfo{title}{{{P}yruvate dehydrogenase and pyruvate dehydrogenase
  kinase expression in non small cell lung cancer and tumor-associated
  stroma}}.
\newblock \emph{\bibinfo{journal}{Neoplasia}} \textbf{\bibinfo{volume}{7}},
  \bibinfo{pages}{1--6} (\bibinfo{year}{2005}).

\bibitem{pmid21300765}
\bibinfo{author}{Vegran, F.}, \bibinfo{author}{Boidot, R.},
  \bibinfo{author}{Michiels, C.}, \bibinfo{author}{Sonveaux, P.} \&
  \bibinfo{author}{Feron, O.}
\newblock \bibinfo{title}{{{L}actate influx through the endothelial cell
  monocarboxylate transporter {M}{C}{T}1 supports an {N}{F}-ë¼{B}/{I}{L}-8
  pathway that drives tumor angiogenesis}}.
\newblock \emph{\bibinfo{journal}{Cancer Res.}} \textbf{\bibinfo{volume}{71}},
  \bibinfo{pages}{2550--2560} (\bibinfo{year}{2011}).

\bibitem{Sonveaux:2008p3954}
\bibinfo{author}{Sonveaux, P.} \emph{et~al.}
\newblock \bibinfo{title}{Targeting lactate-fueled respiration selectively
  kills hypoxic tumor cells in mice}.
\newblock \emph{\bibinfo{journal}{J. Clin. Invest.}}
  \textbf{\bibinfo{volume}{118}}, \bibinfo{pages}{3930--42}
  (\bibinfo{year}{2008}).

\bibitem{pmid21300172}
\bibinfo{author}{Martinez-Outschoorn, U.~E.} \emph{et~al.}
\newblock \bibinfo{title}{{{S}tromal-epithelial metabolic coupling in cancer:
  integrating autophagy and metabolism in the tumor microenvironment}}.
\newblock \emph{\bibinfo{journal}{Int. J. Biochem. Cell Biol.}}
  \textbf{\bibinfo{volume}{43}}, \bibinfo{pages}{1045--1051}
  (\bibinfo{year}{2011}).

\bibitem{pmid24597899}
\bibinfo{author}{Sanita, P.} \emph{et~al.}
\newblock \bibinfo{title}{{{T}umor-stroma metabolic relationship based on
  lactate shuttle can sustain prostate cancer progression}}.
\newblock \emph{\bibinfo{journal}{BMC Cancer}} \textbf{\bibinfo{volume}{14}},
  \bibinfo{pages}{154} (\bibinfo{year}{2014}).

\bibitem{pmid22850421}
\bibinfo{author}{Fiaschi, T.} \emph{et~al.}
\newblock \bibinfo{title}{{{R}eciprocal metabolic reprogramming through lactate
  shuttle coordinately influences tumor-stroma interplay}}.
\newblock \emph{\bibinfo{journal}{Cancer Res.}} \textbf{\bibinfo{volume}{72}},
  \bibinfo{pages}{5130--5140} (\bibinfo{year}{2012}).

\bibitem{pmid24886074}
\bibinfo{author}{Pertega-Gomes, N.} \emph{et~al.}
\newblock \bibinfo{title}{{{A} lactate shuttle system between tumour and
  stromal cells is associated with poor prognosis in prostate cancer}}.
\newblock \emph{\bibinfo{journal}{BMC Cancer}} \textbf{\bibinfo{volume}{14}},
  \bibinfo{pages}{352} (\bibinfo{year}{2014}).

\bibitem{WhitakerMenezes:2011p4016}
\bibinfo{author}{Whitaker-Menezes, D.} \emph{et~al.}
\newblock \bibinfo{title}{Evidence for a stromal-epithelial ``lactate shuttle"
  in human tumors: {MCT4} is a marker of oxidative stress in cancer-associated
  fibroblasts}.
\newblock \emph{\bibinfo{journal}{Cell Cycle}} \textbf{\bibinfo{volume}{10}},
  \bibinfo{pages}{1772--83} (\bibinfo{year}{2011}).

\bibitem{pmid21704401}
\bibinfo{author}{Busk, M.} \emph{et~al.}
\newblock \bibinfo{title}{{{I}nhibition of tumor lactate oxidation:
  consequences for the tumor microenvironment}}.
\newblock \emph{\bibinfo{journal}{Radiother Oncol}}
  \textbf{\bibinfo{volume}{99}}, \bibinfo{pages}{404--411}
  (\bibinfo{year}{2011}).

\bibitem{ardeshir}
\bibinfo{author}{Kianercy, A.}, \bibinfo{author}{Veltri, R.} \&
  \bibinfo{author}{Pienta, K.~J.}
\newblock \bibinfo{title}{{{C}ritical transitions in a game theoretic model of
  tumour metabolism }}.
\newblock \emph{\bibinfo{journal}{Interface Focus}}
  \textbf{\bibinfo{volume}{4}}, \bibinfo{pages}{2014014}
  (\bibinfo{year}{2014}).

\bibitem{pmid21778829}
\bibinfo{author}{Martinez-Outschoorn, U.~E.} \emph{et~al.}
\newblock \bibinfo{title}{{{C}ancer cells metabolically "fertilize" the tumor
  microenvironment with hydrogen peroxide, driving the {W}arburg effect:
  implications for {P}{E}{T} imaging of human tumors}}.
\newblock \emph{\bibinfo{journal}{Cell Cycle}} \textbf{\bibinfo{volume}{10}},
  \bibinfo{pages}{2504--2520} (\bibinfo{year}{2011}).

\bibitem{methods}
\bibinfo{author}{Byrne, M.~B.}, \bibinfo{author}{Leslie, M.~T.},
  \bibinfo{author}{Gaskinsemail, H.~R.} \& \bibinfo{author}{Kenis, P. J.~A.}
\newblock \bibinfo{title}{{{M}ethods to study the tumor microenvironment under
  controlled oxygen conditions}}.
\newblock \emph{\bibinfo{journal}{Trends Biotechnol}}
  \textbf{\bibinfo{volume}{32}}, \bibinfo{pages}{556} (\bibinfo{year}{2014}).

\bibitem{Lemuth:2008p4288}
\bibinfo{author}{Lemuth, K.} \emph{et~al.}
\newblock \bibinfo{title}{Global transcription and metabolic flux analysis of
  {Escherichia} coli in glucose-limited fed-batch cultivations}.
\newblock \emph{\bibinfo{journal}{Appl Environ Microbiol}}
  \textbf{\bibinfo{volume}{74}}, \bibinfo{pages}{7002--15}
  (\bibinfo{year}{2008}).

\bibitem{DeRisi:1997p4290}
\bibinfo{author}{DeRisi, J.~L.}, \bibinfo{author}{Iyer, V.~R.} \&
  \bibinfo{author}{Brown, P.~O.}
\newblock \bibinfo{title}{Exploring the metabolic and genetic control of gene
  expression on a genomic scale}.
\newblock \emph{\bibinfo{journal}{Science}} \textbf{\bibinfo{volume}{278}},
  \bibinfo{pages}{680--6} (\bibinfo{year}{1997}).

\bibitem{Cruz:2012p4421}
\bibinfo{author}{Cruz, R. S. D.~O.} \emph{et~al.}
\newblock \bibinfo{title}{Intracellular shuttle: the lactate aerobic
  metabolism}.
\newblock \emph{\bibinfo{journal}{The Scientific World Journal}}
  \textbf{\bibinfo{volume}{2012}}, \bibinfo{pages}{420984}
  (\bibinfo{year}{2012}).

\bibitem{Jose:2011p4422}
\bibinfo{author}{Jose, C.}, \bibinfo{author}{Bellance, N.} \&
  \bibinfo{author}{Rossignol, R.}
\newblock \bibinfo{title}{Choosing between glycolysis and oxidative
  phosphorylation: a tumor's dilemma?}
\newblock \emph{\bibinfo{journal}{Biochim Biophys Acta}}
  \textbf{\bibinfo{volume}{1807}}, \bibinfo{pages}{552--61}
  (\bibinfo{year}{2011}).

\bibitem{Thiele:2013p4010}
\bibinfo{author}{Thiele, I.} \emph{et~al.}
\newblock \bibinfo{title}{A community-driven global reconstruction of human
  metabolism}.
\newblock \emph{\bibinfo{journal}{Nat Biotechnol}}
  \textbf{\bibinfo{volume}{31}}, \bibinfo{pages}{419--25}
  (\bibinfo{year}{2013}).

\bibitem{DeMartino:2013p4115}
\bibinfo{author}{{{De Martino}}, D.}, \bibinfo{author}{Capuani, F.},
  \bibinfo{author}{Mori, M.}, \bibinfo{author}{{{De Martino}}, A.} \&
  \bibinfo{author}{Marinari, E.}
\newblock \bibinfo{title}{Counting and correcting thermodynamically infeasible
  flux cycles in genome-scale metabolic networks}.
\newblock \emph{\bibinfo{journal}{Metabolites}} \textbf{\bibinfo{volume}{3}},
  \bibinfo{pages}{946--966} (\bibinfo{year}{2013}).

\bibitem{Schellenberger:2012p4302}
\bibinfo{author}{Schellenberger, J.} \emph{et~al.}
\newblock \bibinfo{title}{Predicting outcomes of steady-state
  {$\mathrm{^{13}C}$} isotope tracing experiments using monte carlo sampling}.
\newblock \emph{\bibinfo{journal}{BMC Syst Biol}} \textbf{\bibinfo{volume}{6}},
  \bibinfo{pages}{9} (\bibinfo{year}{2012}).

\bibitem{Price:2004p4298}
\bibinfo{author}{Price, N.~D.}, \bibinfo{author}{Schellenberger, J.} \&
  \bibinfo{author}{Palsson, B.~O.}
\newblock \bibinfo{title}{Uniform sampling of steady-state flux spaces: means
  to design experiments and to interpret enzymopathies}.
\newblock \emph{\bibinfo{journal}{Biophys J}} \textbf{\bibinfo{volume}{87}},
  \bibinfo{pages}{2172--86} (\bibinfo{year}{2004}).

\bibitem{Smith:1996p4127}
\bibinfo{author}{Smith, R.}
\newblock \bibinfo{title}{Efficient Monte Carlo procedures for generating points uniformly distributed over bounded regions}.
\newblock \emph{\bibinfo{journal}{Operations Research}}
  \bibinfo{pages}{1296--1308} (\bibinfo{year}{1984}).

\bibitem{Turcin:1971}
\bibinfo{author}{Turcin, V.}
\newblock \bibinfo{title}{On the computation of multidimensional integrals by
  the {Monte} {Carlo} method}.
\newblock \emph{\bibinfo{journal}{Th Probab Appl}}
  \textbf{\bibinfo{volume}{16}}, \bibinfo{pages}{720--724}
  (\bibinfo{year}{1971}).

\bibitem{Motzkin:1954p4304}
\bibinfo{author}{Motzkin, T.} \& \bibinfo{author}{Schoenberg, I.}
\newblock \bibinfo{title}{The relaxation method for linear inequalities}.
\newblock \emph{\bibinfo{journal}{Canadian J Math}}
  \textbf{\bibinfo{volume}{6}}, \bibinfo{pages}{393--404}
  (\bibinfo{year}{1954}).

\bibitem{Marsaglia:1964p4307}
\bibinfo{author}{Marsaglia, G.} \& \bibinfo{author}{Bray, T.~A.}
\newblock \bibinfo{title}{A convenient method for generating normal variables}.
\newblock \emph{\bibinfo{journal}{SIAM Rev}} \textbf{\bibinfo{volume}{6}},
  \bibinfo{pages}{260--264} (\bibinfo{year}{1964}).

\bibitem{Lovasz:1999p4121}
\bibinfo{author}{Lov{\'a}sz, L.}
\newblock \bibinfo{title}{Hit-and-run mixes fast}.
\newblock \emph{\bibinfo{journal}{Math Program}} \textbf{\bibinfo{volume}{86}},
  \bibinfo{pages}{443--461} (\bibinfo{year}{1999}).

\bibitem{Almaas:2004p4295}
\bibinfo{author}{Almaas, E.}, \bibinfo{author}{Kov{\'a}cs, B.},
  \bibinfo{author}{Vicsek, T.}, \bibinfo{author}{Oltvai, Z.~N.} \&
  \bibinfo{author}{Barab{\'a}si, A.-L.}
\newblock \bibinfo{title}{Global organization of metabolic fluxes in the
  bacterium {Escherichia} coli}.
\newblock \emph{\bibinfo{journal}{Nature}} \textbf{\bibinfo{volume}{427}},
  \bibinfo{pages}{839} (\bibinfo{year}{2004}).

\bibitem{Braunstein:2008p4296}
\bibinfo{author}{Braunstein, A.}, \bibinfo{author}{Mulet, R.} \&
  \bibinfo{author}{Pagnani, A.}
\newblock \bibinfo{title}{Estimating the size of the solution space of
  metabolic networks}.
\newblock \emph{\bibinfo{journal}{BMC Bioinformatics}}
  \textbf{\bibinfo{volume}{9}}, \bibinfo{pages}{240} (\bibinfo{year}{2008}).

\bibitem{Martelli:2009p3766}
\bibinfo{author}{Martelli, C.}, \bibinfo{author}{{De Martino}, A.},
  \bibinfo{author}{Marinari, E.}, \bibinfo{author}{Marsili, M.} \&
  \bibinfo{author}{{P{\'e}rez Castillo}, I.}
\newblock \bibinfo{title}{Identifying essential genes in {Escherichia} coli
  from a metabolic optimization principle}.
\newblock \emph{\bibinfo{journal}{Proc Natl Acad Sci USA}}
  \textbf{\bibinfo{volume}{106}}, \bibinfo{pages}{2607--11}
  (\bibinfo{year}{2009}).

\bibitem{lovazsbook}
\bibinfo{author}{Lov{\'a}sz, L.}
\newblock in \emph{\bibinfo{title}{An algorithmic theory of numbers, graphs and
  convexity}}, \bibinfo{Chapter}{Ch. 2}, \bibinfo{pages}{41--63} ,  (\bibinfo{publisher}{SIAM}, \bibinfo{year}{1984}).

\bibitem{DeMartino:2015p4294}
\bibinfo{author}{{De Martino}, D.}, \bibinfo{author}{Mori, M.} \&
  \bibinfo{author}{Parisi, V.}
\newblock \bibinfo{title}{Uniform sampling of steady states in metabolic
  networks: heterogeneous scales and rounding}.
\newblock \emph{\bibinfo{journal}{PLoS ONE}} \textbf{\bibinfo{volume}{10}},
  \bibinfo{pages}{e0122670} (\bibinfo{year}{2015}).

\end{thebibliography}
\end{document}